%% file: main.tex
\pdfoutput=1
\documentclass[a4paper,10pt]{article}

\usepackage{fullpage}
\usepackage[utf8]{inputenc} 
\usepackage[T1]{fontenc}    
\usepackage[hidelinks]{hyperref}       
\usepackage{url}            
\usepackage{booktabs}       
\usepackage{amsfonts}       
\usepackage{nicefrac}       
\usepackage{microtype}      
\usepackage{xcolor}         
\usepackage{Definitions}
\usepackage{algcompatible}
\usepackage{graphicx}
\usepackage{enumerate}
\usepackage{subfig}
\usepackage{comment}
\usepackage[export]{adjustbox}
\usepackage{mathtools}
\usepackage{caption}
\usepackage{dsfont}
\usepackage{comment}
\usepackage[normalem]{ulem}
\usepackage{authblk}
\usepackage[sort&compress,numbers]{natbib}
\usepackage{amsmath}

\title{Matchings, Predictions and Counterfactual Harm in \\ Refugee Resettlement Processes}

\date{\vspace{-10mm}}

\author[2]{Seungeon Lee\footnote{The author contributed to this paper during an internship at the Max Planck Institute for Software Systems.}} 
\author[1,3]{Nina Corvelo Benz} 
\author[1]{Suhas Thejaswi} 
\author[1]{Manuel Gomez-Rodriguez}

\affil[1]{Max Planck Institute for Software Systems \texttt{\{ninacobe,thejaswi,manuelgr\}@mpi-sws.org}}
\affil[2]{KAIST, \texttt{archon159@kaist.ac.kr}}
\affil[3]{ETH Zurich}

\begin{document}

\maketitle

\begin{abstract}
\input{000abstract}
\end{abstract}

\section{Introduction}
\label{sec:introduction}
\input{010introduction}

\section{A Causal Model for Refugee Resettlement Processes}
\label{sec:scm}
\input{020scm.tex}

\section{On the Optimality of Algorithmic Matching}
\label{sec:optimality}
\input{030optimality}

\section{Algorithmic Matching with Counterfactual Guarantees}
\label{sec:problem}
\input{040problem}

\section{A Practical Post-Processing Framework}
\label{sec:algorithm}
\input{050algorithm}

\section{Experiments}
\label{sec:experiments}
\input{060experiments}

\section{Conclusions}
\label{sec:conclusions}
\input{070conclusions}

\vspace{1mm} \xhdr{Acknowledgements} We would like to thank Evimaria Terzi for fruitful discussions at an early stage of the project. Gomez-Rodriguez acknowledges support from the European Research Council (ERC) under the European Union'{}s Horizon 2020 research and innovation programme (grant agreement No. 945719).

{ 
\bibliographystyle{unsrt}
\bibliography{refs}
}

\clearpage
\newpage

\appendix
\label{sec:appendix}
\input{080appendix}

\end{document}

%% file: 000abstract.tex
Resettlement agencies have started to adopt data-driven algorithmic matching to match refugees to locations using employment rate as a measure of utility.
Given a pool of refugees, data-driven algorithmic matching utilizes a classifier 
to predict the probability that each refugee would find employment at any given location.
Then, it uses the predicted probabilities to estimate the expected utility of all possible 
placement decisions.
Finally, it finds the placement decisions that maximize the predicted utility by solving a maximum weight bipartite matching problem.
In this work, we argue that, using existing solutions, there may be pools of refugees for which data-driven algorithmic matching is (counterfactually) harmful---it would have achieved lower utility than a 
given default policy used in the past, had it been used.
Then, we develop a post-processing algorithm that, given placement decisions made by a default policy on a pool of refugees and their employment outcomes,
solves an inverse~matching problem to minimally modify the predictions made by a given classifier. Under these modified predictions, the optimal matching policy that maximizes predicted utility 
on the pool is guaranteed to be not harmful.
Further, we introduce a Transformer model that, given placement decisions made by a default policy on multiple pools of refugees and their employment outcomes, learns to modify the predictions made by a classifier so that the optimal matching policy that maximizes predicted utility under the modified predictions on an unseen pool of refugees is less likely to be harmful than under the original predictions.
Experiments on simulated resettlement processes using synthetic refugee data created from publicly available sources suggest that our methodology may be effective in making algorithmic placement decisions that are less likely to be harmful than existing solutions.

%% file: 010introduction.tex
In recent years, there is an increasing excitement in the potential of data-driven algorithmic matching to improve matching decisions in a wide variety of high-stakes application domains. 
Examples of such~ma\-tching~decisions include: 
matching refugees to locations~\cite{bansak2018improving, ahani2021placement,ahani2023dynamic,freund2023group};
matching patients to~appoint\-ments in health clinics~\cite{salah2022predict}; or
matching blood/organ donations to recipients~\cite{mcelfresh2023matching,aziz2021optimal}.
In all these cases, a central authority needs to distribute a limited set of resources---locations, appointments, or donations---among a group of individuals---refugees, patients, or recipients.

In this work, we focus on data-driven algorithmic matching in refugee resettlement processes,
a specific application domain where matching decisions impact a particularly vulnerable group of individuals, data-driven algorithmic matching has been already deployed by a large resettlement agency in the United States\footnote{\href{https://www.refugees.ai/}{https://www.refugees.ai/}}, 
and may be soon deployed by others in the United States and elsewhere\footnote{\href{https://rematch-eu.org/about-matching/}{https://rematch-eu.org/about-matching/}}.
Given a pool of refugees, data-driven algorithmic matching aims to optimize the overall utility of placement decisions for these refugees. 
Because employment has been argued to play an important factor in the success of integration, e.g., by enabling self-sufficiency and ties to local residents~\cite{ager2008understanding}, this utility is typically measured as the number of refugees that find employment soon
after relocation.
If we knew beforehand in which locations refugees would find employment, 
then we could find the optimal decision policy by solving a maximum weight bipartite matching problem~\cite{tanimoto1978some, lau2011iterative}.
In this matching problem, nodes would represent refugees as well as locations and edge weights would represent employment outcomes, \ie, whether each refugee would ($1$) or would not ($0$) find employment at each location, and the goal would be to find a matching that maximizes the sum of edge weights in the matching.
Unfortunately, at the time placement decisions are made, we cannot know for sure in which locations refugees would find employment---there is uncertainty on the value of the employment outcomes.
To overcome this challenge, previous work on data-driven algorithmic matching leverages machine learning classifiers to predict the probability that a refugee would find employment at each location and then uses these predicted probabilities as edge weights in the above matching problem~\cite{bansak2018improving, ahani2021placement,ahani2023dynamic,freund2023group}.
As a result, given a pool of refugees, the resulting algorithmic decision policy is guaranteed to maximize predicted utility---the expected utility over the predicted employment probabilities.

However, previous work on data-driven algorithmic matching does not only utilize predicted utility to make algorithmic placement decisions but also to evaluate the quality of these decisions in com\-pa\-ri\-son with
decisions made by a default policy used in the past.
In doing so, they implicitly assume that, for each location, the predicted employment probabilities are well-calibrated estimates of the true employment pro\-ba\-bi\-li\-ties~\cite{gneiting2007probabilistic}. 
Unfortunately, such an assumption is likely to be violated because every placement decision policy induces a different distribution of refugees across locations---it induces a different distribution shift~\cite{quinonero2008dataset}. 
As a consequence, using existing solutions, any claim of superiority of algorithmic placement decisions over decisions made by default policy used in the past based on predicted utility is questionable.
In fact, one cannot even tell to what extent algorithmic matching implements the principle of ``first, do no harm'', a principle that has been recently argued to be applicable to machine learning systems for decision support~\cite{richens2022counterfactual,beckers2022causal,li2023trustworthy,beckers2023quantifying}.\footnote{The \href{https://www.europarl.europa.eu/doceo/document/TA-9-2024-0138-FNL-COR01_EN.pdf}{European Unions'{} AI} act mentions the term ``harm'' more than $35$ times and points out that, its crucial role in the design of algorithmic systems must be defined carefully.}  

In this context, the most widely accepted definition of harm is arguably the counterfactual comparative account of harm (in short, counterfactual harm)~\cite{feinberg1986wrongful,hanser2008metaphysics,klocksiem2012defense}.
Under this definition, an action is harmful to~an~in\-di\-vi\-dual if they would~have been in a worse state had the action been taken.
Building upon this definition, in our work, we say that data-driven algorithmic matching is harmful to a pool of refugees if it would have worsened their employment outcomes in comparison with a default decision policy used in the past, had it been used.
Then, our goal is to minimize how frequently data-driven algorithmic matching causes harm.

\xhdr{Our Contributions}
%
%
We start by formally characterizing resettlement processes in terms of a structural causal model (SCM)~\citep{pearl2009causality}, as illustrated in Figure~\ref{fig:scm}.
Using this characterization, 
%
%
%
we first show that, 
if the decisions made by an algorithmic decision policy satisfy a counterfactual condition with respect to the placement decisions made by a given default policy on a pool of refugees, 
then the algorithmic decision policy is not harmful to this pool---it would have achieved equal or higher utility than the default policy, had it been used.
Building on this counterfactual condition, we make the following contributions:
%
\squishlist
    \item[(1)] We develop an algorithm that, given placement decisions made by a default policy on a pool of refugees and the (corresponding) realized employment outcomes, 
    solves an inverse bipartite matching problem to minimally modify the refugees' predicted employment probabilities provided by a given classifier. 
    Under these modified probabilities, the placement decisions made by an algorithmic matching policy that maximizes the predicted utility provably satisfy the above counterfactual condition on the pool.
    
    \item[(2)] Given placement decisions made by a default policy on multiple pools of refugees and the (corresponding) realized employment outcomes,
    we use the predicted employment probabilities provided by a classifier and the minimally modified predicted employment probabilities provided by the previously introduced algorithm to train a Transformer model. 
    This model is able to minimally modify the predicted employment probabilities of an unseen pool so that it approximately satisfies the above counterfactual condition with respect to this pool.
\squishend
%
Finally, we validate our methodological contributions on simulated resettlement processes using synthetic refugee data created using publicly available aggregated data from a variety of international organizations, including the United Nations Refugee Agency (UNHCR). 
The results show that our methodology may be effective in making algorithmic placement decisions that are less likely to be harmful than existing solutions\footnote{We provide the code and data used in our experiments in following github link: \url{https://github.com/Networks-Learning/counterfactually-harmless-matching}}.

\xhdr{Further related work}
Bipartite matching problems, or more generally assignment problems, have multifaceted applications in various domains and thus a rich, extensive literature \cite{gale1962college, crawford1981job, lovasz2009matching, gibbons1985algorithmic, stelmakh2021peer, mcelfresh2023matching, aziz2021optimal}. 
Less common and not well known are inverse assignment problems~\cite{demange2014introduction, heuberger2004inverse, lee2020inverse}. 
Whereas assignment problems aim to find an optimal assignment that maximizes a desired parameterized objective function, inverse assignment problems aim to find the minimal change to those parameters such that a desired assignment becomes an optimal assignment~\cite{liu2003inverse,berczi2023inverse}.
Inverse assignment problems have been recently proven useful in the context of counterfactual explanations~\cite{korikov2021counterfactual, korikov2021counterfactualconstraint}. In particular, finding the nearest counterfactual explanation for algorithmic assignments has been reduced to an inverse problem.
Within the literature on inverse assignment problems, 
the work most closely related to ours is
by Yang and Zhang~\cite{yang2007partial}, which introduces an algorithmic framework to solve partial inverse assignment problems 
where the desired assignment is only partially given.
More specifically, in our first methodological contribution, we adapt their framework 
to mi\-ni\-mally mo\-di\-fy the refugees' predicted employment probabilities by a given classifier (in their framework, the parameters) so that, 
under these modified probabilities, 
the optimal placement decisions (the optimal assignment) match a subset of those made by a default policy (the desired partial assignment).

%% file: 020scm.tex
We consider a resettlement process where, for each realization of the process, 
a decision maker receives a pool $\Ical$ of $n$ refugees with features $\xb = (x_i)_{i \in \Ical} \in \Xcal^{n}$, 
matches each refugee $i \in \Ical$ to a location $\lb = (l_i)_{i \in \Ical} \in \Lcal^{n}$ out of $k$ locations, 
and eventually receives a utility $u(\yb) \geq 0$, where $\yb = (y_i)_{i \in \Ical} \in \{0, 1\}^{n}$ are outcome variables specifying whether a refugee $i$ finds a job ($y_i = 1$) or does not find a job ($y_i=0$) soon after relocation.
Here, we assume that each location $l \in \Lcal$ has a maximum capacity $c_l$ to host refugees and the utility $u(\yb) = \sum_{i} u(y_i)$ is separable. 
Without loss of generality, we further assume that $u(y_i)=y_i$.

\begin{figure}[t]
\centering
\includegraphics[width=.6\textwidth]{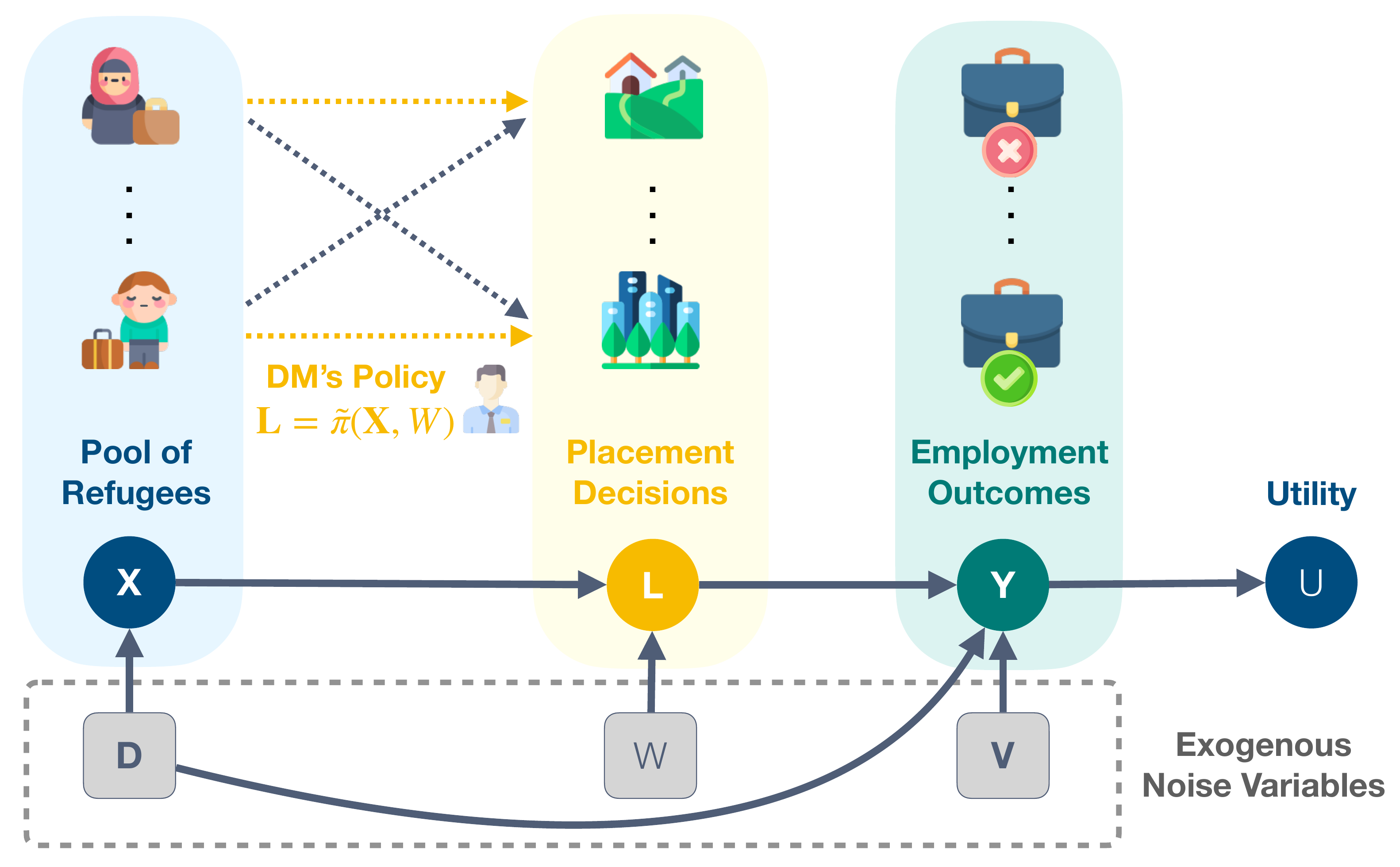}
\caption{Our structural causal model $\Mcal$. Circles represent endogenous random variables and boxes represent exogenous random variables. The value of each endogenous variable is given by a function of the values of its ancestors in the structural causal model, as defined by Eq.~\ref{eq:scm}. The value of each exogenous variable is sampled independently from a given distribution.}
\label{fig:scm}
\end{figure}

Next, we characterize the matching process using a structural causal model (SCM)~\cite{pearl2009causality}, which we denote as $\Mcal$. 
The SCM $\Mcal$, which entails a distribution $P^{\Mcal}$, is defined by the following set of assignments\footnote{Random variables are denoted with capital letters and realizations of random variables with lower case letters.}:
\begin{equation} \label{eq:scm}
    X_i = f_{X}(D_i), \quad Y_i = f_{Y}(D_i, V_{i, L_i}) \quad \forall i \in \Ical, \quad 
    \Lb = \tilde{\pi}(\Xb, W) \quad  \text{and} \quad U = \mathbf{1}^{T} \Yb,
\end{equation}
where $D_i \sim P(D)$, $V_{i,l} \sim P(V \given L=l)$ and $W \sim P(W)$ are independent exogenous random variables, often called exogenous noise variables, which characterize 
the refugee's individual characteristics\footnote{To allow for features $X_i$ that are causal and anticausal to the outcome variable $Y_i$, the noise variable $D_i$ is a parent of both $X_i$ and $Y_i$, as discussed elsewhere~\cite{scholkopf2012causal}. If we allow only for causal features and there are no hidden confounders, we could just write $Y_i = f_{Y}(X_i, V_{i, L_i})$.}, 
the synergies between locations and refugees, 
and 
the decision maker's individual~cha\-rac\-te\-ris\-tics, respectively.
Further, we have that $f_{X}$ and $f_Y$ are unknown causal functions, 
and $\tilde{\pi}$ is the decision maker's default matching policy (in short, the default
policy)\footnote{For ease of presentation, we assume a constant pool size $n$ and capacities $c_l$.
However, all theoretical results and algorithms can be easily adapted to settings where the pool
size and capacities change across matching processes.}.
Figure~\ref{fig:scm} shows a visual representation of our SCM $\Mcal$.

Moreover, we assume the decision maker has access to a classifier $g : \Lcal \times \Xcal \rightarrow [0, 1]^{|\Lcal|}$ that, 
for each location $l \in \Lcal$, 
maps a refugee'{}s feature vector $x \in \Xcal$ to a 
%
%
predicted employment probability $g_{l}(x)$. 
Here, the higher the predicted employment probability $g_l(x)$, the more the classifier $g$ believes the candidate will find a job at location $l$, and we denote the predicted probabilities for all refugees in the pool $\Ical$ as $\gb=(g_{l}(x_i) )_{i \in \Ical, l \in \Lcal}$.
Further, let $\Pi(\Gb)$ be the class of algorithmic matching policies that assign refugees to locations based on the predicted employment probabilities $\Gb$ and satisfy the capacity constraints.
Then, we can view the implementation of an algorithmic matching policy $\pi(\Gb) \in \Pi(\Gb)$ as an intervention $\text{do}(\Lb = \pi(\Gb))$ in the SCM $\Mcal$.
The intervened SCM $\Mcal^{\text{do}(\Lb = \pi(\Gb))}$, which entails a distribution $P^{\Mcal \,;\, \text{do}(\Lb = \pi(\Gb))}$, is defined by: 
\begin{equation} \label{eq:intervened-scm} 
X_i = f_{X}(D_i), \,\, Y_i = f_{Y}(D_i, V_{i, L_i}), \,\, G_{i,l} = g_{l}(X_i) \,\, \forall i \in \Ical, \forall l \in \Lcal, \,\,
\Lb = \pi(\Gb) \,\, \text{and} \,\, U = \mathbf{1}^{T} \Yb.
\end{equation}
In addition, given placement decisions $\lb$ made by a default policy $\tilde{\pi}$ on a pool of refugees with features $\xb$ and the (corresponding) realized employment outcomes $\yb$, 
we define a counterfactual SCM $\Mcal_{\Xb = \xb, \Yb = \yb, \Lb = \lb}$ where the noise variables are distributed according to the posterior distribution $P(\Db, \Vb, W\given \Xb = \xb, \Yb = \yb, \Lb = \lb)$ and we denote the resulting distribution entailed by this SCM using $P^{\Mcal \given \Xb = \xb, \Yb = \yb, \Lb = \lb}$.
Finally, we characterize counterfactual statements comprising the algorithmic policy $\pi(\Gb)$ using the intervened counterfactual SCM $\Mcal^{\text{do}(\Lb = \pi(\gb))}_{\Xb = \xb, \Yb = \yb, \Lb = \lb}$ and denote its entailed distribution using $P^{\Mcal \given \Xb = \xb, \Yb = \yb, \Lb = \lb \,;\, \text{do}(\Lb = \pi(\gb))}$.

In the next section, we will use the above characterization to reason about the conditions the predicted employment probabilities $\Gb$ should satisfy so that the optimal matching policy that maximizes predicted utility also maximizes the expected utility $\EE[U\given \Xb]$.

%% file: 030optimality.tex
Given a pool $\Ical$ of $n$ refugees with features $\xb$, 
existing approaches to data-driven algorithmic matching equate the problem of finding an optimal algorithmic matching policy that maximizes (conditional) expected utility, \ie,
\begin{equation} \label{eq:expected-utility-optimization}
\pi^{\ast}(\gb) \in \Pi^{\ast}(\gb) = \argmax_{\pi(\gb) \in \Pi(\gb)} \, \EE_{\Yb \sim P^{\Mcal \,;\, \text{do}(\Lb = \pi(\gb))}}[\mathbf{1}^{T} \Yb \given \Xb = \xb],
\end{equation}
to the much easier problem of finding an algorithmic matching policy that maximizes (conditional) predicted utility, \ie, 
\begin{equation} \label{eq:predicted-utility-optimization}
\hat{\pi}(\gb) \in \hat{\Pi}(\gb) = \argmax_{\pi(\gb) \in \Pi(\gb)} \, \sum_{i \in \Ical} g_{\pi_{i}(\gb)}(x_i),
\end{equation}
where $\pi_{i}(\gb) \in \Lcal$ denotes the location assigned by policy $\pi(\gb)$ to refugee $i \in \Ical$.
This latter problem can be viewed as a maximum weight bipartite matching problem~\cite{tanimoto1978some, lau2011iterative} and thus its solution, which may be non unique, can be recovered from the solution to the following linear program:
\begin{equation} \label{eq:matching-lp}
\begin{split}
  \text{maximize} \quad & \quad \sum_{i \in \Ical, l \in \Lcal} g_{l}(x_i) \, z_{il} \\
    \text{subject to} \quad & \quad \sum_{l \in \Lcal} z_{il} \leq 1 \quad \forall i \in \Ical, \\
    \quad & \quad \sum_{i \in \Ical} z_{il} \leq c_l \quad \forall l \in \Lcal, \\
    \quad & \quad z_{il} \in \{0, 1\} \quad \forall i \in \Ical, l \in \Lcal.
\end{split}
\end{equation}
In particular, it holds that $\hat{\pi}(\gb) = (\argmax_{l \in \Lcal} \hat{z}_{il})_{i \in \Ical}$, where $\hat{\zb} = (\hat{z}_{il})_{i \in \Ical, l \in \Lcal}$ is an optimal integral solution to the above linear program, as shown elsewhere~\cite{bansak2018improving, ahani2021placement}. 
%
However, existing approa\-ches~do not really investigate the sufficient conditions under which the predicted employment probabilities $\gb$ should~sa\-tis\-fy for both policies $\pi^{\ast}(\gb)$ and $\hat{\pi}(\gb)$ to offer the same expected utility.

The following proposition shows that, if the predicted employment probabilities $\gb$ are perfectly calibrated, then policies in $\hat{\Pi}(\gb)$ and $\Pi^{\ast}(\gb)$ must offer the same expected utility.\footnote{All proofs can be found in Appendix~\ref{app:proofs}.}
\begin{proposition}
\label{prop:optimality}
For any $\xb \sim P^{\Mcal}(\Xb)$, if 
$g_{l}(x_i) = P^{\Mcal \,;\, do(L_i =
  l)}(Y_i = 1 \given X_i = x_i)$ for all $l \in \Lcal$ and $i \in \Ical$,
then, for any $\hat{\pi}(\gb) \in \hat{\Pi}(\gb)$, it holds that $\hat{\pi}(\gb) \in \Pi^{\ast}(\gb)$.
%
%
\end{proposition}
Unfortunately, the above condition is unlikely to hold in practice.
This is due to the fact that, if the distribution of conditional probability values 
$P^{\Mcal \,;\, do(\Lb = \lb)}(Y_i = 1 \given X_i)$ induced by $P^{\Mcal}(X_i)$ is nonatomic, even
if we are able to sample data from $P^{\Mcal \,;\, do(\Lb = \lb)}$, finding the individual
probabilities $P^{\Mcal \,;\, do(L_i = l)}(Y_i = 1 \given X_i = x_i)$ from this data is not possible without distributional assumptions, even asymptotically~\cite{barber2020distribution,gupta2020distribution}.
Consequently, we cannot say whether the use of existing approaches to data-driven algorithmic matching over a given default policy for a particular pool of refugees will truly increase utility or cause potential harm. With this in mind, over the next two sections, we introduce an alternative approach to data-driven algorithmic matching with the goal of reducing harmful placement decisions with respect to the default policy.
%
%
%
%
%
%
%
%

%

%% file: 040problem.tex
%
%
%
%

Given the placement decisions $\lb$ made by a default policy $\tilde{\pi}$ on a pool of refugees $\Ical$ with features $\xb$ and the (corresponding) realized employment outcomes $\yb$,
we first identify a class of algorithmic decision policies that are counterfactually harmless---they would have achieved at least the same utility as the default policy, had they been used:
\begin{proposition} \label{prop:counterfactual-harm} 
Given $\xb, \yb, \lb \sim P^{\Mcal}(\Xb, \Yb, \Lb)$, let the class of algorithmic decision policies $\Pi_{\xb,\yb,\lb}$ be defined as
\begin{equation} \label{eq:non-harmful-policies}
\Pi_{\xb,\yb,\lb} = \{ \pi(\gb) \given \pi(\gb) \in \Pi(\gb) \wedge \pi_i(\gb) = l_i \,\, \forall i \in \Ical \,\, \text{such that} \,\, y_i = 1 \}.
\end{equation}
Then, for any $\pi(\gb) \in \Pi_{\xb,\yb,\lb}$, it holds that 
\begin{equation} \label{eq:counterfactual-harm}
    \mathbf{1}^{T} \Yb \geq \mathbf{1}^{T}\yb \quad \text{for all} \quad \Yb \sim P^{\Mcal \given \Xb = \xb, \Yb = \yb, \Lb = \lb \,;\, \text{do}(\Lb = \pi(\gb))}(\Yb),
\end{equation}
where $P^{\Mcal \given \Xb = \xb, \Yb = \yb, \Lb = \lb \,;\, \text{do}(\Lb = \pi(\gb))}(\Yb)$ 
denotes the intervened counterfactual distribution of the employment outcomes $\Yb$ entailed by the counterfactual SCM $\Mcal^{\text{do}(\Lb = \pi(\gb))}_{\Xb = \xb, \Yb = \yb, \Lb = \lb}$.
\end{proposition}
Here, it is important to note that
the expected utility achieved by any given algorithmic policy with respect to all possible pools of refugees
can be rewritten as an average over counterfactual utilities, as formalized by the following proposition:
\begin{proposition} \label{prop:counterfactual-equality}
For any resettlement process $\Mcal$ satisfying Eq.~\ref{eq:scm} and algorithmic policy $\pi \in \Pi(\Gb)$, 
the following equality holds:
\begin{equation} \label{eq:cfc_rewriten_exp}
    \EE_{\Yb \sim P^{\Mcal \,;\, \text{do}(\Lb = \pi(\Gb))}(\Yb)}[\mathbf{1}^{T} \Yb] 
    = \EE_{\Xb', \Yb', \Lb' \sim P^{\Mcal}(\Xb',\Yb',\Lb')} \left[ \EE_{\Yb \sim P^{\Mcal \given \Xb = \Xb', \Yb = \Yb', \Lb = \Lb' \,;\, \text{do}(\Lb = \pi(\Gb))}(\Yb)}[\mathbf{1}^{T} \Yb] \right]
\end{equation}
\end{proposition}
Then, from Propositions~\ref{prop:counterfactual-harm} and~\ref{prop:counterfactual-equality}, we can immediately conclude that,
if an algorithmic policy $\pi$ is counterfactually harmless for \emph{any} pool of refugees $\Ical$, 
then, $\pi$ will achieve equal or higher expected utility than the default policy $\tilde{\pi}$, 
\ie, if $\pi(\gb) \in \Pi_{\xb, \yb, \lb}$ for any $\xb, \yb, \lb \sim P^{\Mcal}(\Xb, \Yb, \Lb)$, then, it holds that
%
%
%
\begin{equation} \label{eq:harmless-average}
    \EE_{\Yb \sim P^{\Mcal \,;\, \text{do}(\Lb = \pi(\Gb))}(\Yb)}[\mathbf{1}^{T} \Yb] \geq \EE_{\Yb \sim P^{\Mcal}(\Yb)}[\mathbf{1}^{T} \Yb].
\end{equation}

In the next section, we will build upon the above theoretical results to develop a practical post-pro\-cessing framework to minimally modify the predicted employment probabilities $\gb=g(\xb)$ provided by a given classifier $g$
so that the algorithmic matching policy $\hat{\pi}(\breve{\gb})$ that maximizes (conditional) 
predicted utility under the minimally modified probabilities $\breve{\gb}$ is less likely to 
be counterfactually harmful than the algorithmic matching policy $\hat{\pi}(\gb)$ that maximizes (conditional) predicted utility under the original probabilities $\gb$.

%% file: 050algorithm.tex
%
Given retrospective data about multiple pools of refugees under the default policy $\tilde{\pi}$, our framework first finds, for each of the above pools, the minimally modified predicted probabilities $\breve{\gb}$ under which any $\hat{\pi}(\breve{\gb}) \in \hat{\Pi}(\breve{\gb})$ is provably counterfactually harmless with respect to $\tilde{\pi}$, in hindsight.
Then, it uses the original and modified predicted probabilities of all the above pools to train a transformer model $h$ that, given the original predicted probabilities $\gb$ of an 
\emph{unseen} pool $\Ical$,
predicts the minimally modified predicted probabilities $\breve{\gb}$.\footnote{Our framework allows for the modified predicted probabilities $\breve{\gb}$ to be larger than $1$.}

\xhdr{Avoiding Counterfactual Harm, In Hindsight} 
%
Given $\xb, \yb, \lb \sim P^{\Mcal}(\Xb, \Yb, \Lb)$, we formulate the problem of finding the minimally modified predicted employment probabilities $\breve{\gb}$ under which any $\hat{\pi}(\breve{\gb}) \in \hat{\Pi}(\breve{\gb})$ is provably counterfactually harmless with respect to $\tilde{\pi}$ as follows:
\begin{equation} \label{eq:partial_inverse}
    \begin{split}
    \text{minimize}  &\quad ||\breve{\gb} - \gb||_1 = \sum_{i \in \Ical, l \in \Lcal} |\breve{g}_l(x_i) - g_l(x_i)|, \\
    \text{subject to} & \quad \hat{\Pi}(\breve{\gb}) \subseteq \Pi_{\xb, \yb, \lb}.
    \end{split}
\end{equation}
To solve the above problem, we build on the algorithmic framework of Yang and Zhang~\cite{yang2007partial}. 
Let $\pi'(\gb) \in \Pi_{\xb, \yb, \lb}$ be a policy that maximizes the (conditional) predicted utility over the refugees $i \in \Ical' = \{ i \in \Ical \given l_i = 0\}$ and note that,
for any given $\gb$, $\pi'(\gb)$ can be recovered from the solution to the following linear program, similarly as in Eqs.~\ref{eq:predicted-utility-optimization} and~\ref{eq:matching-lp}:
\begin{equation} \label{eq:partial-lp}
    \begin{split}
  \text{maximize} \quad & \quad \sum_{i \in \Ical', l \in \Lcal} g_{l}(x_i) \, z_{il} \\
    \text{subject to} \quad & \quad \sum_{l \in \Lcal} z_{il} \leq 1 \quad \forall i \in \Ical', \\
    \quad & \quad \sum_{i \in \Ical'} z_{il} \leq c'_{l} \quad \forall l \in \Lcal, \\
    \quad & \quad z_{il} \in \{0, 1\} \quad \forall i \in \Ical', l \in \Lcal,
\end{split}
\end{equation}
where $c'_{l} = c_l - |\{i \in \Ical \given l_i = l \wedge y_i = 1\}|$. In particular, we have that $\pi'_i(\gb) = \argmax_{l \in \Lcal} z'_{il}$ if $i \in \Ical'$, where $\zb' = (z'_{il})_{i \in \Ical', l \in \Lcal}$ is an optimal integral 
solution to the above linear program, and $\pi'_i(\gb) = l_i$ otherwise. 

Then, 
using the above policy $\pi'(\gb)$ and the optimal solution to the dual of the linear program 
defined by Eq.~\ref{eq:matching-lp},
%
%
we find the minimally modified predicted employment probabilities $\breve{\gb}$, as formalized by the following Theorem:
\begin{theorem} \label{thm:minimally-modified-predicted-probabilities}
    Given $\xb, \yb, \lb \sim P^{\Mcal}(\Xb, \Yb, \Lb)$, the minimally modified employment probabilities $\breve{\gb}$ under which any algorithmic policy $\hat{\pi}(\breve{\gb}) \in \hat{\Pi}(\breve{\gb})$ is provably counterfactually harmless with respect to the default policy $\tilde{\pi}$, as defined in Eq.~\ref{eq:partial_inverse}, is given by:
    \begin{equation} \label{eq:breve-gb}
        \breve{g}_l(x_i) = \left\{
        \begin{array}{l@{}l}
        \hat{u}_i + \hat{v}_l + \epsilon & \quad \text{if} \,\,\, \pi'_i(\gb) = l  \\
        g_{l}(x_i) & \quad \text{otherwise},
    \end{array}
    \right.
    \end{equation}
    where
    $\epsilon > 0$ is an infinitesimally small constant 
    and
    $\hat{\ub} = (\hat{u}_i)_{i \in \Ical}$ and $\hat{\vb} = (\hat{v}_l)_{l \in \Lcal}$ are the optimal solution to the dual of the linear program defined by Eq.~\ref{eq:matching-lp} under predicted probabilities $\gb$, \ie,
    \begin{equation} \label{eq:partial-dual}
    \begin{split}
    \text{minimize}  &\quad \sum_{i \in \Ical} u_i + \sum_{l \in \Lcal} c_{l} v_{l}, \\
    \text{subject to} & \quad u_i + v_l \geq g_{l}(x_i) \quad \forall i \in \Ical, l \in \Lcal, \\
    & \quad u_i \geq 0, v_{l} \geq 0 \quad \forall i \in \Ical, l \in \Lcal.
    \end{split}
    \end{equation}
\end{theorem}

\xhdr{Learning to Avoid Counterfactual Harm} 
%
In the previous section, we have derived an expression to compute the minimally modified employment 
probabilities $\breve{\gb}$ under which any $\hat{\pi}(\breve{\gb}) \in \hat{\Pi}(\breve{\gb})$ 
is provably counterfactually harmless with respect to the default policy $\tilde{\pi}$.
Unfortunately, the expression requires retrospective data about the pool of interest, \ie, the placement decisions $\lb$ made by a default policy $\tilde{\pi}$ and the (corresponding) realized employment
outcomes $\yb$. 
As a consequence, we cannot directly use it to compute the minimally mo\-di\-fied employment probabilities $\breve{\gb}$ of unseen pools.

However, we can use the above expression on retrospective data from multiple pools of refugees $\{ \Ical_j \}$
to train a deep learning model $h : [0, 1]^{k \times n} \rightarrow \RR_{+}^{k \times n}$ that, given the original predicted probabi\-li\-ties $\gb$ of an unseen pool $\Ical$, approximately predicts the minimally modified predicted probabilities $h(\gb) \approx \breve{\gb}$.
Our model consists of four modules: 
(i) a prediction probability projection layer,
(ii) a capacity projection layer,
(iii) $N$ layers of Transformer encoders,
and (iv) an embedding projection layer.
The projection la\-yers comprise two linear projections with ReLU activation.
The layers of Transformer encoders exclude positional encoding so that they become agnostic to the number or order of the refugees 
in each pool.
Moreover, in our experiments, we train our model using a quadratic loss, \ie, $\EE_{\xb, \yb, \lb\sim P^{\Mcal}(\Xb, \Yb, \Lb)} \left[ \sum_{i \in \Ical, l \in \Lcal} \left( h_{l,i}(\gb) - \breve{g}_{l}(x_i) \right)^{2} \right]$, where $h_{l, i}(\gb)$ denotes the approximately minimally modified predicted probability for $l \in \Lcal$ and $i \in \Ical$.
For further implementation details used in our experiments, refer to Appendix~\ref{app:additional-implementation-details}.

%% file: 060experiments.tex
In this section, we use publicly available aggregated data from a variety of international organizations, including the United Nations Refugee Agency (UNHCR), to generate synthetic refugee data. 
Then, we use this synthetic data to simulate and compare the outcome of multiple resettlement processes under a default policy and several algorithmic policies.
Here, note that, by using synthetic refugee data, rather than real refugee data, 
we can calculate the (true) expected utility achieved by any algorithmic policy 
and, given retrospective data under a default policy,
we can calculate the (true) counterfactual utility achieved by any algorithmic policy.\footnote{We are not aware of any publicly available dataset with real refugee data. Unfortunately, we were unable to obtain access to real refugee data used in previous studies~\cite{bansak2018improving, ahani2021placement,ahani2023dynamic,freund2023group}.}

\begin{table}[t]
\centering
\caption{Percentage (\%) of pools counterfactually harmed by the algorithmic policies $\hat{\pi}(\pb)$, $\hat{\pi}(\gb)$, and $\hat{\pi}(h(\gb))$ in the test set under various noise levels $w$ and $\beta = 0.6$. 
Lower numbers indicate better performance.
The policy $\hat{\pi}(\breve{\gb})$ consistently achieves 0\% harm, as expected from Eq.~\ref{eq:harmless-average}, and thus it is not presented. 
Bold numbers indicate that $\hat{\pi}(h(\gb))$ counterfactually harms fewer pools than $\hat{\pi}(\gb)$, with results averaged over 5 runs. More quantitative results, including standard deviations, are available in Appendix~\ref{app:additional-experiment-results}.
}
\vspace{1mm}
\begin{tabular}{@{}lccccccccc@{}}
\toprule
Noise Level ($w$) & 0 & 0.125 & 0.25 & 0.375 & 0.5 & 0.625 & 0.75 & 0.875 & 1 \\ \midrule
$\hat{\pi}(\pb)$ & 0 & 33.2 & 23.6 & 17.6 & 13.6 & 11.4 & 9.2 & 8.2 & 6 \\
$\hat{\pi}(\gb)$ & 75 & 64.6 & 55 & 48.2 & 41 & 33.4 & 24.8 & 23.4 & 15.6 \\
$\hat{\pi}(h(\gb))$ & \textbf{69.2} & \textbf{58.9} & \textbf{45.4} & \textbf{38.0} & \textbf{33.9} & \textbf{30.6} & \textbf{22.9} & 23.4 & 16.0 \\ \bottomrule
\end{tabular}
\label{tab:harm_pool_ratio}
\end{table}

\xhdr{Experimental Settings}
We create $5{,}000$ synthetic pools of refugees to be resettled to $k = 10$ locations. 
Each pool $\Ical$ contains $n=100$ synthetic refugees, and each refugee $i \in \Ical$ is represented by $4$-dimensional feature vector $x_i$ and $k=10$ labels $\{ y_i(l) \}_{l \in \Lcal}$ indicating whether the refugee would find employment (or not) at each state $l$ soon after relocation.
Each location corresponds to one US State and the features contain demographic information about a refugee'{}s age, country of origin, sex, and educational attainment.
The features and labels are sampled from distributions $P^{\Mcal}(X)$ and $P^{\Mcal \,;\, \text{do}(L_i=l)}(Y \given X=x_i)$ informed by aggregate statistics from UNHCR, the World Bank, the U.S. Census, the U.S. Bureau of Labor Statistics, and Migration Policy Institute. 
Refer to Appendix~\ref{app:synthetic-data-generation} for more details on the synthetic data generation process.
The classifier $g$ overestimates the value of the true employment probability for half of the locations, picked at random, and underestimates its value for the remaining half. 
Whenever $g$ overestimates the employment probability, it predicts $g_{l}(x) = p \cdot (1 + \beta)$ and, whenever $g$ underestimates the employment probability, it predicts $g_{l}(x) = p \cdot (1 - \beta)$,
where $p = P^{\Mcal \,;\, \text{do}(L_i=l)}(Y_i=1 \given X_i=x_i)$ and $\beta$ is a given parameter.
%
%
%
\begin{figure}[t]
    \centering
    \subfloat[Low noise level, $w=0.25$]{
        \includegraphics[width=.32\linewidth]{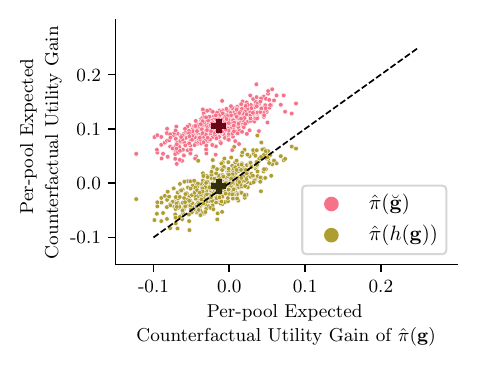}
    }
    \subfloat[Medium noise level, $w=0.5$]{
        \includegraphics[width=.32\linewidth]{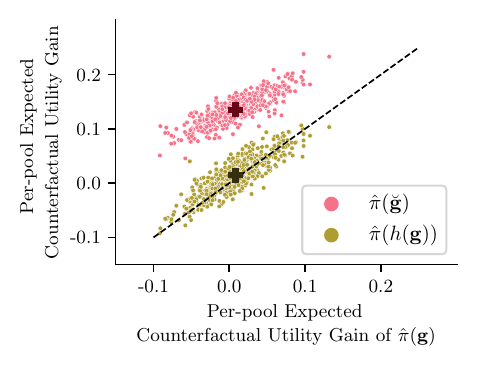}
    }
    \subfloat[High noise level, $w=0.875$]{
        \includegraphics[width=.32\linewidth]{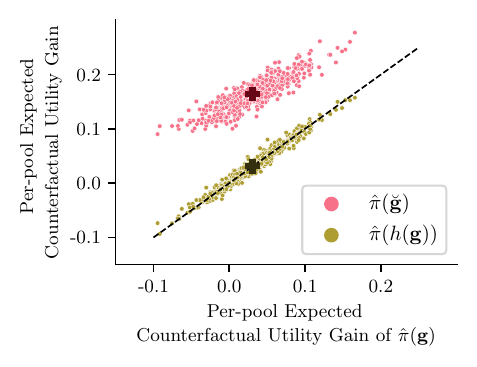}
    }
    \vspace{-1mm}
    \caption{
    Per-pool expected counterfactual utility gain achieved by the proposed algorithmic policies $\hat{\pi}(\breve{\gb})$ and $\hat{\pi}(h(\gb))$ with respect to the policy $\hat{\pi}(\gb)$ in the test set for $\beta = 0.6$ under low, medium and high noise level. 
    The cross markers indicate the expected counterfactual utility gain across all pools in the test set. 
    Pools above the (dashed) identity line (\ie, $y=x$) show increase in counterfactual utility gain compared to policy $\hat{\pi}(\gb)$.
    %
    %
    }
    \label{fig:distribution-utility-gain}
    \vspace{-3mm}
\end{figure}

In our experiments, we randomly split the pools into a training set with $4{,}000$ pools, which we use to train the deep learning model $h$, 
a validation set with $500$ pools, which we use for searching the best model across multiple epochs as well as the best value of $\epsilon$ for each default policy $\tilde{\pi}(\xb, w)$,
and a test set with $500$ pools, which we use for evaluation\footnote{Refer to Appendix~\ref{app:additional-implementation-details} for additional implementation details regarding the deep learning model $h$.}.
Moreover, we implement and compare the performance of the following policies:
%
\squishlist
%

\item[(i)] A default policy $\tilde{\pi}(\xb, w)$ that first finds the placement decisions that maximize the (conditional) expected utility under the true employment probabilities $\pb$ 
and then picks a ratio $w$ of these placement decisions and shuffles them with each other.

\item[(ii)] An algorithmic policy $\hat{\pi}(\pb)$ that makes placement decisions that maximize the (conditional) expected utility under the true employment probabilities $\pb$. 
%
This policy is unreali\-za\-ble in practice, as discussed in Section~\ref{sec:optimality}.
Here, note that $\hat{\pi}(\pb)=\tilde{\pi}(\xb, 0)$.

\item[(iii)] An algorithmic policy $\hat{\pi}(\gb)$ that makes placement decisions that maximize the (conditional) predicted utility under the predicted employment probabilities $\gb$. 

\item[(iv)] An algorithmic policy $\hat{\pi}(\breve{\gb})$ that makes placement decisions that maximize the (conditional) predicted utility under the modi\-fied predicted employment probabilities $\breve{\gb}$ given by Eq.~\ref{eq:breve-gb}.
This policy is unrealizable in practice since it uses retrospective data about the pool of interest. 

\item[(v)] An algorithmic policy $\hat{\pi}(h(\gb))$ that makes placement decisions that maximize the (conditional) predicted utility under the  postprocessed predicted employment probabilities $h(\gb)$, where $h$ is the deep learning model described in Section~\ref{sec:algorithm}.
\squishend
%
To compare the performance achieved by the above policies, we use 
the percentage of pools in the test set that are counterfactually harmed by each algorithmic policy $\hat{\pi}(\cdot)$
and 
the expected counterfactual utility $\EE_{\Yb \sim P^{\Mcal \given \Xb = \xb, \Yb = \yb, \Lb = \lb \,;\, \text{do}(\Lb = \hat{\pi}(\cdot))}(\Yb)}[\mathbf{1}^{T} \Yb]$ achieved by each algorithmic policy $\hat{\pi}(\cdot)$ in comparison with the realized utility achieved by the default policy $\tilde{\pi}(\xb, w)$ across pools of refugees in the test set.
%
%
%
In Appendix~\ref{app:additional-experiment-results}, we also report the (realized) utility achieved by each policy across pools of refugees in the test set.
%
\begin{figure}[t]
    \centering
    \subfloat[Low noise level, $w=0.25$]{
        \includegraphics[width=.32\linewidth]{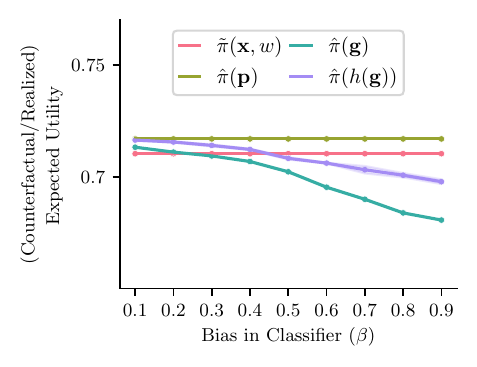}
    }
    \subfloat[Medium noise level, $w=0.5$]{
        \includegraphics[width=.32\linewidth]{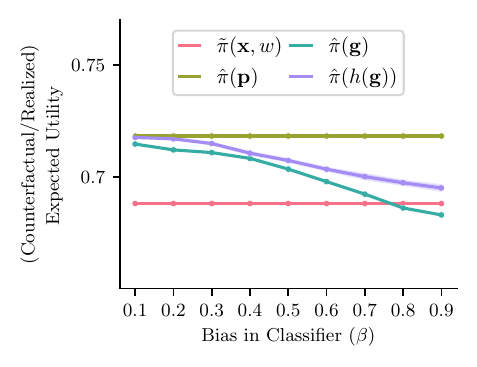}
    }
    \subfloat[High noise level, $w=0.875$]{
        \includegraphics[width=.32\linewidth]{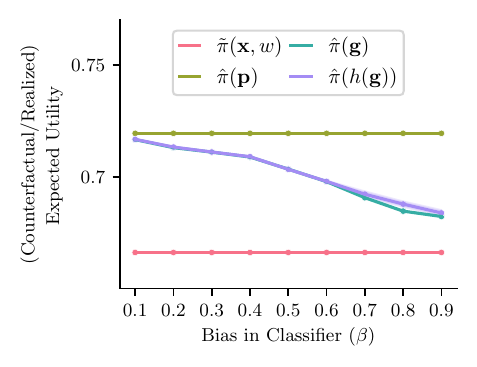}
    }
    \vspace{-1mm}
    \caption{
    Expected counterfactual utility achieved by the algorithmic policies $\hat{\pi}(\pb)$, $\hat{\pi}(\gb)$ and $\hat{\pi}(h(\gb))$ in comparison with the expected realized utility achieved by the default policy $\tilde{\pi}(\xb, w)$ across all pools in the test set for different $\beta$ values under low, medium and high noise level.
    For $\hat{\pi}(h(\gb))$, the results are averaged over $5$ runs, where the error bands represent standard deviations. 
    %
    }
    \label{fig:utility-gain-by-beta-1}
    \vspace{-4mm}
\end{figure}

\xhdr{Results} We first calculate the percentage of pools in the test set that are counterfactually harmed by each algorithmic policy under different noise levels $w$.
Table~\ref{tab:harm_pool_ratio} presents the results for $\beta = 0.6$, which offer several insights. 
%
%
We find that, by postprocessing the predicted employment probabilities $\gb$, the algorithmic policy $\hat{\pi}(h(\gb))$ counterfactually harms fewer (similar) pools than the algorithmic policy $\hat{\pi}(\gb)$ for low (high) noise levels, \ie, $w \leq 0.75$ ($w > 0.75$).
This suggest that, as the level of noise of the default policy $\tilde{\pi}(\xb, w)$ increases, it is more difficult for our framework to learn to avoid harm from past placement decisions made by the default policy and the (corresponding) realized employment outcomes.

Next, we compare the per-pool expected counterfactual utility gain achieved by the algorithmic policies $\hat{\pi}(\breve{\gb})$ and $\hat{\pi}(h(\gb))$, which are designed using our framework, in comparison to 
the counterfactual utility gain achieved by algorithmic policy $\hat{\pi}(\gb)$ under low ($w=0.25$), 
medium ($w=0.5$) and high ($w=0.75$) noise level.
Figure~\ref{fig:distribution-utility-gain} summarizes the results for $\beta = 0.6$, which further supports the findings derived from Table~\ref{tab:harm_pool_ratio}.
As expected from Proposition~\ref{prop:counterfactual-harm}, we find that the unreali\-za\-ble algorithmic policy $\hat{\pi}(\breve{\gb})$ always achieves a positive expected counterfactual gain for every pool.
Further, for low and medium noise levels, we find that the algorithmic policy $\hat{\pi}(h(\gb))$ typically offers a greater per-pool expected counterfactual utility gain than the algorithmic policy $\hat{\pi}(\gb)$
and it achieves a positive expected counterfactual utility averaged across pools.
For high noise levels, we find that both algorithmic policies $\hat{\pi}(h(\gb))$ and $\hat{\pi}(\gb)$ offer a comparable expected counterfactual utility gain per pool.

Finally, we compare the expected counterfactual utility achieved by the algorithmic policies $\hat{\pi}(\breve{\gb})$ and $\hat{\pi}(h(\gb))$ in comparison with the expected realized utility achieved by the default policy $\tilde{\pi}(\xb, w)$ for different $\beta$ values under low ($w=0.25$), 
medium ($w=0.5$) and high ($w=0.75$) noise level.
Figure~\ref{fig:utility-gain-by-beta-1} summarizes the results, which show that, for low and medium noise levels, the algorithmic policy $\hat{\pi}(h(\gb))$ offers greater expected counterfactual utility than $\hat{\pi}(\gb)$ across all $\beta$ values and, for high noise levels, both offer a comparable expected counterfactual utility.

%% file: 070conclusions.tex
In this work, we have initiated the study of (counterfactual) harm in data-driven algorithmic mat\-ching.
We have developed a postprocessing framework that, given retrospective data under a given default policy, 
postprocesses the predictions made by the classifiers used in existing solutions to avoid causing harm.
Further, using synthetic refugee data created using a variety of publicly available data, we have empirically shown that our framework may be effective in making algorithmic placement decisions that are less likely to be harmful than existing solutions.
Our work opens up many interesting avenues for future work. 
For example, our work considers a setting in which data-driven algorithmic matching replaces a human decision maker.
However, it would be interesting to consider a setting in which data-driven algorithmic matching supports, rather than replace, the human decision maker.
Moreover, it would be important to evaluate our methodological contributions on real retrospective refugee data comprising of placement decisions and realized employment 
outcomes.
Further, we have focused on reducing the overall amount of harm caused by data-driven algorithmic matching. 
However, this may lead significant disparities across demographic groups, as shown recently~\cite{freund2023group}, and thus it would be important to extend our methodology to account for fairness considerations.
Finally, while we have focused on refugee resettlement processes, our ideas may also be proven useful in other domains such as matching patients to appointments in health clinics~\cite{salah2022predict}, matching blood/organ donors to recipients~\cite{mcelfresh2023matching,aziz2021optimal}, or  matching reviewers to submissions in conferences~\cite{saveski2023counterfactual}.

%% file: 080appendix.tex
\section{Proofs} \label{app:proofs} 
\input{081optimality}
\input{082inverse}

\clearpage
\newpage
\section{Additional Details about the Synthetic Data Generation}
\label{app:synthetic-data-generation}
\input{083synthetic}

\clearpage
\newpage
\section{Additional Implementation Details} \label{app:additional-implementation-details}
\input{084implementation}

\clearpage
\newpage
\section{Additional Experimental Results}
\label{app:additional-experiment-results}
\input{085additional_experiments}

%% file: 081optimality.tex
\subsection{Proof of Proposition~\ref{prop:optimality}}
We prove this by contra-position. For a given $\Xb=\xb$, if the algorithmic matching policy $\hat{\pi}(\gb)=(\argmax_{l \in \Lcal} \hat{z}_{il})_{i \in \Ical}$, where $\hat{\zb} = (\hat{z}_{il})_{i \in \Ical, l \in \Lcal}$ is an optimal integral solution to the linear program defined by Eq.~\ref{eq:matching-lp}, does not correspond to an optimal policy $\pi^{\ast}(\gb)$, then we must have that 
\begin{equation} \label{eq:non_optimal}
\EE_{\Yb \sim P^{\Mcal \,;\, \text{do}(\Lb = \pi^{\ast}(\gb))}}[\mathbf{1}^{T} \Yb \given \Xb = \xb] > \EE_{\Yb \sim P^{\Mcal \,;\, \text{do}(\Lb = \hat{\pi}(\gb))}}[\mathbf{1}^{T} \Yb \given \Xb = \xb].
\end{equation}
%

Given any algorithmic policy $\pi(\gb)$, let $\zb = (z_{il})_{i \in \Ical, l \in \Lcal}$ with $z_{il} = \mathds{1}[\pi_i(\gb) = l]$, where note that $z_{il}$ has to satisfy the constraints of the matching problem defined in Eq.~\ref{eq:matching-lp} for $\pi(\gb)$ to be a valid algorithmic matching policy.
That is, each refugee is assigned to at most one location and each location $l$ is assigned at most $c_l$ refugees.
Further, we have that: 
\begin{equation*}
\begin{split}
        \EE_{\Yb \sim P^{\Mcal \,;\, \text{do}(\Lb = \pi(\gb))}}[\mathbf{1}^{T} \Yb \given \Xb = \xb] 
        &=\sum_{i\in \Ical} P^{\Mcal \,;\, do(L_i = (\pi_i(\gb)))}(Y_i = 1 \given X_i = x_i) \\
        &= \sum_{i\in \Ical, l \in \Lcal}  P^{\Mcal \,;\, do(L_i = l)}(Y_i = 1 \given X_i = x_i) \mathds{1}[\pi_i(g(\xb)) = l] \\
        &= \sum_{i\in \Ical, l \in \Lcal}  g_l(x_i)  z_{il}
\end{split} 
\end{equation*}
Hence, we can rewrite Eq.~\ref{eq:non_optimal} as
\begin{equation*}
    \sum_{i\in \Ical, l \in \Lcal}  g_l(x_i)  z^\ast_{il} > \sum_{i\in \Ical, l \in \Lcal}  g_l(x_i)  \hat{z}_{il}
\end{equation*}
where $z^\ast_{il}= \mathds{1}[\pi^\ast_i(\gb)=l]$.
However, this leads to a contradiction since the solution $\hat{\zb}$ cannot be the optimal solution to the linear program defined by Eq.~\ref{eq:matching-lp} if $\zb^\ast$ is a feasible solution and has greater objective function value. 
As a consequence, it must hold that 
\begin{equation*} 
\EE_{\Yb \sim P^{\Mcal \,;\, \text{do}(\Lb = \pi^{\ast}(\gb))}}[\mathbf{1}^{T} \Yb \given \Xb = \xb] \leq \EE_{\Yb \sim P^{\Mcal \,;\, \text{do}(\Lb = \hat{\pi}(\gb))}}[\mathbf{1}^{T} \Yb \given \Xb = \xb]
\end{equation*}
and thus $\hat{\pi}(\gb) \in \Pi^{\ast}(\gb)$.

\subsection{Proof of Proposition~\ref{prop:counterfactual-harm}}

We begin by writing the counterfactual probability of $Y_i = 1$, for $i \in \Ical$, in terms of the probability density of $D_i$, $ V_{i,l_i}$ and $ V_{i,\pi_i(\gb)}$ as an expectation:
\begin{multline} \label{eq:cfc_exp}
P^{\Mcal \given \Xb = \xb, \Yb = \yb, \Lb = \lb \,;\, \text{do}(\Lb = \pi(\gb))}(Y_i=1) 
        =\EE_{Y_i \sim P^{\Mcal \given \Xb = \xb, \Yb = \yb, \Lb = \lb \,;\, \text{do}(\Lb = \pi(\gb))}}[Y_i] \\
        = \int_{d, v, v'} P(D_i=d, V_{i,l_i}=v, V_{i,\pi_i(\gb)}=v' \given \Xb=\xb, \Yb=\yb, \Lb=\lb) \cdot f_Y(d,v')\,dd\,dv \,dv'.
\end{multline}
Since $V_{i,\pi_i(\gb)}$ is independent of $D_i, X_i$ and $Y_i$ conditioned on $V_{i,l_i}$, we can rewrite the probability density function in Eq.~\ref{eq:cfc_exp} as
\begin{multline} \label{eq:cfc_density}
        P(D_i=d, V_{i,l_i}=v \given X_i=x_i, Y_i=y_i, L_i=l_i) \cdot P(V_{i,\pi_i(\gb)}=v' \given V_{i,l_i}=v)\\
        =P(D_i=d, V_{i,l_i}=v \given f_X(D_i)=x_i, f_Y(D_i, V_{i,l_i})=y_i) \cdot P(V_{i,\pi_i(\gb)}=v' \given V_{i,l_i}=v),
\end{multline}
Further, it is straight forward to see that
\begin{equation} \label{eq:cfc_v}
    P(V_{i,\pi_i(\gb)}=v' | V_{i,l_i}=v) = \begin{cases}
        1 & \text{if } \pi_i(\gb) = l_i \text{ and } v'=v \\
        0 & \text{if } \pi_i(\gb) = l_i \text{ and } v'\neq v \\
        P(V_{i,\pi_i(\gb)}=v') = P(V=v' \given L_i=\pi_i(\gb) ) & \text{if } \pi_i(\gb) \neq l_i ,
    \end{cases}
\end{equation}
where the last case follows from the definition of SCM $\Mcal$ since $V_{i,l_i}$ and $V_{i,\pi_i(\gb)}$ are independent when $\pi_i(\gb) \neq l_i$.
Thus, when $\pi_i(\gb) = l_i$, it follows from Eqs.~\ref{eq:cfc_density} and \ref{eq:cfc_v} that the right hand side of Eq.~\ref{eq:cfc_exp} is equivalent to
\begin{multline}\label{eq:equality_y}
         \int_{d, v} P(D_i=d, V_{i,l_i}=v \given f_X(D_i)=x_i, f_Y(D_i, V_{i,l_i})=y_i) \cdot f_Y(d,v)\,dd\,dv \\
         = y_i \cdot \int_{d, v} P(D_i=d, V_{i,l_i}=v \given f_X(D_i)=x_i, f_Y(D_i, V_{i,l_i})=y_i) \,dd\,dv = y_i,
\end{multline}
where the equality stems from the fact that density function $P(D_i=d, V_{i,l_i}=v \given f_X(D_i)=x_i, f_Y(D_i, V_{i,l_i})=y_i)$ can only be positive when $f_Y(d,v)=y_i$.
Since $y_i \in \{0,1\}$, it follows from Eq.~\ref{eq:equality_y} that for $\Yb \sim P^{\Mcal \given \Xb = \xb, \Yb = \yb, \Lb = \lb \,;\, \text{do}(\Lb = \pi(\gb))}$ must hold that
\begin{equation}\label{eq:equality_Y}
    Y_i=y_i \quad \text{ for all } i\in \Ical \text{ such that } \pi_i(\gb)=l_i.
\end{equation}

Now, let $\pi(\gb)$ be a policy in $\Pi_{\xb,\yb,\lb}$\,.
By definition of $\Pi_{\xb,\yb,\lb}$, we have that, for any $i\in \Ical$, $y_i=1$ implies $\pi_i(\gb) = l_i$. Hence, we can conclude that
\begin{equation*}
\mathbf{1}^{T}\yb = \sum_{i \in \Ical} y_i
= \sum_{i \in \Ical} y_i \cdot \mathds{1}[\pi_i(\gb) = l_i]
\overset{(i)}{=}  \sum_{i \in \Ical} Y_i \cdot \mathds{1}[\pi_i(\gb) = l_i]
\leq \mathbf{1}^{T}\Yb .
\end{equation*}
where (i) follows from Eq.~\ref{eq:equality_Y}.

\subsection{Proof of Proposition~\ref{prop:counterfactual-equality}}

Using linearity of expectation, we can rewrite the right hand side of Eq.~\ref{eq:cfc_rewriten_exp} as
\begin{multline} \label{eq:cfc_exp_full}
\EE_{\Xb', \Yb', \Lb' \sim P^{\Mcal}} \left[
\EE_{\Yb \sim P^{\Mcal \given \Xb = \Xb', \Yb = \Yb', \Lb = \Lb' \,;\, \text{do}(\Lb = \pi(\Gb))}}[\mathbf{1}^{T} \Yb] \right]
\\= \int_{\xb'} \sum_{\yb'} \sum_{\lb'} P^\Mcal(\Xb'=\xb', \Yb'=\yb', \Lb'=\lb') \left[
    \EE_{\Yb \sim P^{\Mcal \given \Xb = \Xb', \Yb = \Yb', \Lb = \Lb' \,;\, \text{do}(\Lb = \pi(\Gb))}}[\mathbf{1}^{T} \Yb] \right] \, d\xb' 
\end{multline}
Using Eq.~\ref{eq:cfc_exp} and the fact that $\Xb' \overset{d}{=} \Xb$ and $\Yb' \overset{d}{=} \Yb$, we have that
\begin{align}\label{eq:inner_exp}
 \begin{split}
        P^\Mcal&(\Xb'=\xb', \Yb'=\yb', \Lb'=\lb') \left[
    \EE_{\Yb \sim P^{\Mcal \given \Xb = \Xb', \Yb = \Yb', \Lb = \Lb' \,;\, \text{do}(\Lb = \pi(\Gb))}}[\mathbf{1}^{T} \Yb] \right]
    \\= & \int_{\db', \vb', \vb} P^\Mcal(\Lb'=\lb' \given \Xb'=\xb') \times P(\Db=\db', \Vb_{:,\Lb'}=\vb', \Vb_{:,\pi(\Gb)}=\vb, \Xb'=\xb', \Yb'=\yb'\given  \Lb'=\lb') 
      \\& \qquad \times \left[\sum_{i\in \Ical}f_Y(d'_i,v_i)\right]\,d\db'\,d\vb' \,d\vb
 \end{split}
\end{align}
where $\Vb_{:,\Lb'}=(V_{i,L'_i})_{i\in \Ical}$ and $\Vb_{:,\pi(\Gb)}=(V_{i,\pi_i(\Gb)})_{i\in \Ical}$ are vectors of exogeneous noise variables $V_{i,l}$ with $ i \in \Ical, l \in \Lcal$. 

Note that, when combining Eq.~\ref{eq:inner_exp} with Eq.~\ref{eq:cfc_exp_full}, we can omit random variables $\Xb'$ and $\Yb'$ from the joint distribution $P(\Db, \Vb_{:,\Lb'},\Vb_{:,\pi(\Gb)}, \Xb', \Yb' \given \Lb')$ and substitute it from the conditional distribution $P^\Mcal(\Lb'=\lb' \given \Xb'=\xb')$ since the outcome of both variables can be deduced deterministically from exogenous variables $\Db$ and $\Vb_{:,\Lb'}$. We obtain following expression 
\begin{multline}\label{eq:noise_cfc_exp_full}
\EE_{\Xb', \Yb', \Lb' \sim P^{\Mcal}} \left[
\EE_{\Yb \sim P^{\Mcal \given \Xb = \Xb', \Yb = \Yb', \Lb = \Lb' \,;\, \text{do}(\Lb = \pi(\Gb))}}[\mathbf{1}^{T} \Yb] \right]
\\=
    \sum_{\lb'}\int_{\db', \vb', \vb}  P^\Mcal(\Lb'=\lb' \given \Db'=\db') \cdot P(\Db=\db', \Vb_{:,\Lb'}=\vb', \Vb_{:,\pi(\Gb)}=\vb \given \Lb'=\lb') \\ \times \left[\sum_{i\in \Ical}f_Y(d'_i,v_i)\right]\,d\db'\, d\vb'\, d\vb
\end{multline}

Using that $\Db, \Vb_{:,\pi(\Gb)}$ are independent from $\Lb'$
and rearranging summations and integrals in the above equation, we have that
\begin{align*}
\int_{\db', \vb} P(\Db=\db', \Vb_{:,\pi(\Gb)}=\vb)  
     \left[\sum_{i\in \Ical}f_Y(d'_i,v_i) \right]
     &\times \sum_{\lb'} P^\Mcal(\Lb'=\lb' \given \Db'=\db')
     \\ &\times \int_{\vb'} P(\Vb_{:,\Lb'}=\vb'\given \Lb'=\lb', \Vb_{:,\pi(\Gb)}=\vb)\, d\vb'\,d\db' d\vb\\ 
      \\ &\overset{(i)}{=} \int_{d', v} P(\Db=\db') \cdot P(\Vb_{:,\pi(\Gb)}=\vb) \cdot \sum_{i \in \Ical} f_Y(d'_i,v_i)\,d\db' d\vb \label{eq:simplified_inner_exp}
      \\ &\overset{(ii)}{=} \EE_{\Yb \sim P^{\Mcal \,;\, \text{do}(\Lb = \pi(\Gb))}}[\mathbf{1}^{T} \Yb]
\end{align*}
where (i) follows from $ \sum_{\lb'} P^\Mcal(\Lb'=\lb' \given \Db'=\db') = 1$ and the fact that $\int_{\vb'} P(\Vb_{:,\Lb'}=\vb'\given \Lb'=\lb', \Vb_{:,\pi(\Gb)}=\vb)\, d\vb' =1$ and (ii) follows from the definition of intervened SCM $\Mcal^{\text{do}(\Lb = \pi(\Gb))}$.

%% file: 082inverse.tex
\subsection{Proof of Theorem~\ref{thm:minimally-modified-predicted-probabilities}}
We first find the minimally modified predicted probabilities $\gb'$ under which the counterfactually harmless policy $\pi'(\gb) \in \Pi_{\xb, \yb, \lb}$ that maximizes the (conditional) predicted utility over the refugees in $\Ical' = \{i \in \Ical \given l_i = 0\}$ with respect to $\gb$ also maximizes the (conditional) predicted utility over refugees in $\Ical$ with respect to $\gb'$, \ie, 
\begin{equation} \label{eq:inverse}
    \begin{split}
    \text{minimize}  &\quad ||\gb' - \gb||_1 \\
    \text{subject to} & \quad \pi'(\gb) \in \hat{\Pi}(\gb'),
    \end{split}
\end{equation}
where note that, under the minimally modified predicted probabilities $\gb'$, there might exist other $\hat{\pi}(\gb') \in \hat{\Pi}(\gb')$ such that $\hat{\pi}(\gb') \notin \Pi_{\xb, \yb, \lb}$. 
%
In what follows, we assume that the pool size $n = \sum_{l \in \Lcal} c_{l}$  without loss of generality.\footnote{If $n < \sum_{l \in \Lcal} c_l$, 
we can introduce a (dummy) refugee set $\Ical'$ of size $\sum_{l \in \Lcal} c_l - n$ with predicted probabilities $g_{l}(x_i) = 0$ for every $i \in \Ical'$ and $l \in \Lcal$. 
If $\sum_{l \in \Lcal} c_l < n$, we can introduce a (dummy) location $l'$ with capacity 
$c_{l'} = n - \sum_{l \in \Lcal} c_{l}$ with predicted probabilities $g_{l'}(x_i) = 0$ for every $i \in \Ical$. 
In both cases, the optimal solution $\gb'$ does not change.} 

To solve the above problem, we apply the algorithmic framework by Yang and Zhang~\cite{yang2007partial}.
First, we claim that, for all $i \in \Ical$ and $l \in \Lcal$, the minimally modified predicted probability $g'_{l}(x_i)$ must satisfy that
\begin{equation} \label{eq:property-solution-inverse}
\begin{split}
g'_{l}(x_i) &\geq g_{l}(x_i) \,\, \text{if} \,\, \exists \hat{\pi}(\gb') \in \hat{\Pi}(\gb') \,\, \text{such that} \,\, \hat{\pi}_{i}(\gb') = l \\
\quad g'_{l}(x_i) &\leq g_{l}(x_i) \,\, \text{if} \,\, \nexists \hat{\pi}(\gb') \in \hat{\Pi}(\gb') \,\, \text{such that} \,\, \hat{\pi}_{i}(\gb') = l,
\end{split}
\end{equation}

We establish this claim by contradiction. 
Assume there exists $\hat{\pi}(\gb') \in \hat{\Pi}(\gb')$ with at least one $i \in \Ical$ such that 
$\hat{\pi}_i(\gb') = l$ and $g'_{l}(x_i) < g_{l}(x_i)$ .
This contradicts the optimality of $\gb'$ as a solution for Eq.~\ref{eq:inverse}, since maintaining the original predicted probability $g_{l}(x_i)$ yields a better solution.
Assume there exists at least one $i \in \Ical$ such that $\nexists \hat{\pi}(\gb') \in \hat{\Pi}(\gb')$ with $\hat{\pi}_i(\gb') = l$ and $g'_{l}(x_i) > g_{l}(x_i)$. 
This also contradicts the optimality of $\gb'$ as maintaining the original predicted probability $g_{l}(x_i)$ for $i \in \Ical$ also yields a better solution.

Using the above claim, we can derive a lower bound on the objective value achieved by the minimally
modified predicted probabilities $\gb'$, as formalized by the following Lemma:
\begin{lemma}
\label{lemma:lower-bound}
The minimally modified predicted probabilities $\gb'$ under which $\pi'(\gb) \in \hat{\Pi}(\gb')$
satisfy that
\begin{equation*}
|| \gb' - \gb ||_1
\geq \sum_{i \in \Ical} \left[ g_{\hat{\pi}_i(\gb)}(x_i) - g_{\hat{\pi}_i(\gb')}(x_i) \right]
\end{equation*}
for any $\hat{\pi}(\gb) \in \hat{\Pi}(\gb)$ and $\hat{\pi}(\gb') \in \hat{\Pi}(\gb')$.
\end{lemma}
Further, we can use the dual of the linear program used to find an algorithmic matching policy $\hat{\pi}(\gb) \in \hat{\Pi}(\gb)$, defined in Eq.~\ref{eq:partial-dual}, to show that the 
above lower bound is essentially tight, as formalized by the following Proposition:
\begin{proposition} \label{prop:equality}
The minimally modified predicted probabilities $\gb'$ under which $\pi'(\gb) \in \hat{\Pi}(\gb')$ 
satisfy that:
\begin{equation*}
||\gb' - \gb||_1 = \sum_{i \in \Ical} \left[ g_{\hat{\pi}_i(\gb)}(x_i) - g_{\hat{\pi}_i(\gb')}(x_i) \right]
\end{equation*}
for any $\hat{\pi}(\gb) \in \hat{\Pi}(\gb)$ and $\hat{\pi}(\gb') \in \hat{\Pi}(\gb')$.
\end{proposition}
Importantly, the proof of the above proposition directly gives us an explicit solution to the problem of finding the minimally modified predicted probabilities $\gb'$ under which $\pi'(\gb) \in \hat{\Pi}(\gb')$:
\begin{equation*}
        g'_l(x_i) = \left\{
        \begin{array}{l@{}l}
        \hat{u}_i + \hat{v}_l & \quad \text{if} \,\,\, \pi'_i(\gb) = l  \\
        g_{l}(x_i) & \quad \text{otherwise},
        \end{array}
        \right.
\end{equation*}
where $\hat{\ub} = (\hat{u}_i)_{i \in \Ical}$ and $\hat{\vb} = (\hat{v}_l)_{l \in \Lcal}$ is the optimal solution to the dual of the linear program used to find an algorithmic matching policy $\hat{\pi}(\gb) \in \hat{\Pi}(\gb)$.

Further, we can show that $\gb'$ as defined above is also a solution to the problem of finding the minimally modified predicted probabilities such that there exists $\hat{\pi}(\gb') \in \hat{\Pi}(\gb')$ with $\hat{\pi}(\gb') \in 
\Pi_{\xb, \yb, \lb}$. 
This claim is proven by contradiction.
%
Assume there exists $\pi''(\gb) \in \Pi_{\xb, \yb, \lb}$ such that,
for the minimally modified predicted probabilities $\gb''$ such that $\pi''(\gb) \in \hat{\Pi}(\gb'')$,
it holds that $|| \gb'' - \gb ||_1 < || \gb' - \gb ||_1$. Since both $\pi''(\gb)$ and $\pi'(\gb)$ are counterfactually harmless, we have that $\pi''_i(\gb)=\pi'_i(\gb)$ for all $i \in \Ical \setminus \Ical'$. Hence, using Proposition~\ref{prop:equality}, it must follow that 
$\sum_{i \in \Ical'}  g_{\pi'_i(\gb)}(x_i) < \sum_{i \in \Ical'} g_{\pi''_i(\gb)}(x_i)$. This directly contradicts the optimality of $\pi'(\gb)$ as the policy that maximizes the (conditional) predicted utility over refugees in $\Ical'$ described in Eq.~\ref{eq:partial-lp}.

Then, we can immediately conclude that the minimally modified employment probabilities $\breve{\gb}$ under
which any $\hat{\pi}(\breve{\gb}) \in \hat{\Pi}(\breve{\gb})$ is provably counterfactually harmless as defined in Eq.~\ref{eq:partial_inverse} is given by:
\begin{equation*}
        \breve{g}_l(x_i) = \left\{
        \begin{array}{l@{}l}
        g'_l(x_i) + \epsilon & \quad \text{if} \,\,\, \pi'_i(\gb) = l  \\
        g'_l(x_i) & \quad \text{otherwise},
    \end{array}
    \right.
\end{equation*}
where $\epsilon > 0$ is an infinitesimally small constant that rules out the possibility that, under the minimally modified predicted probabilities $\breve{\gb}$, there exists $\hat{\pi}(\breve{\gb}) \in \hat{\Pi}(\breve{\gb})$ such that $\hat{\pi}(\breve{\gb}) \notin \Pi_{\xb, \yb, \lb}$.
This concludes the proof.

\subsection{Proof of Lemma~\ref{lemma:lower-bound}}
For any $\hat{\pi}(\gb) \in \hat{\Pi}(\gb)$ and $\hat{\pi}(\gb') \in \hat{\Pi}(\gb')$, we have that:
\begin{align*}
\sum_{i \in \Ical, l \in \Lcal} |g'_{l}(x_i) - g_{l}(x_i)|
&\geq \sum_{i \in \Ical} \left[ |g'_{\hat{\pi}_i(\gb')}(x_i) - g_{\hat{\pi}_i(\gb')}(x_i)| + \mathds{1}(\hat{\pi}_i(\gb) \neq \hat{\pi}_i(\gb')) |g'_{\hat{\pi}_i(\gb)}(x_i) - g_{\hat{\pi}_i(\gb)}(x_i)| \right]
\\
&\overset{(i)}{=} \sum_{i \in \Ical} \left[ (g'_{\hat{\pi}_i(\gb')}(x_i) - g_{\hat{\pi}_i(\gb')}(x_i))
+ \mathds{1}(\hat{\pi}_i(\gb) \neq \hat{\pi}_i(\gb')) (g_{\hat{\pi}_i(\gb)}(x_i) - g'_{\hat{\pi}_i(\gb)}(x_i)) \right]
\\
&= \sum_{i \in \Ical} \left[ g'_{\hat{\pi}_i(\gb')}(x_i) - g'_{\hat{\pi}_i(\gb)}(x_i) + g_{\hat{\pi}_i(\gb)}(x_i) - g_{\hat{\pi}_i(\gb')}(x_i) \right. \\
& \left. \qquad \qquad \qquad \qquad +\, \mathds{1}(\hat{\pi}_i(\gb) = \hat{\pi}_i(\gb')) (g'_{\hat{\pi}_i(\gb)}(x_i) - g_{\hat{\pi}_i(\gb)}(x_i)) \right]
\\
& \overset{(ii)}{\geq} \sum_{i \in \Ical} \left[ g_{\hat{\pi}_i(\gb)}(x_i) - g_{\hat{\pi}_i(\gb')}(x_i) \right],
\end{align*}
where (i) follows from both inequalities in Eq.~\ref{eq:property-solution-inverse} and (ii) follows from the definition of $\hat{\pi}_i(\gb')$
and
the first inequality in Eq.~\ref{eq:property-solution-inverse}.

\subsection{Proof of Proposition~\ref{prop:equality}}
Let $\hat{z}(\gb)$ denote an optimal solution to the linear program defined by Eq.~\ref{eq:matching-lp} under the predicted probabilities $\gb$, 
and 
$\hat{\ub}(\gb)$ and $\hat{\vb}(\gb)$ denote the optimal solution to the dual of the same linear 
program also under $\gb$.

From the complementary slackness conditions, for every $i \in \Ical$ and $l \in \Lcal$, it 
holds that:
\begin{equation*}
\begin{split}
\hat{u}_i(\gb) + \hat{v}_l(\gb) &= g_{l}(x_i) \quad \text{if} \,\, \hat{\pi}_{i}(\gb) = l  \\
\hat{u}_i(\gb) + \hat{v}_l(\gb) &\geq g_{l}(x_i) \quad \text{otherwise}.
\end{split}
\end{equation*}
Let $\hat{\gb} = \{\hat{g}_{l}(x_i)\}_{i \in \Ical, l \in \Lcal}$ where, for each $i \in \Ical$ and $l \in \Lcal$, we set the value of $\hat{g}_{l}(x_i)$ as follows:
\begin{equation} \label{eq:invlp_weights}
\hat{g}_{l}(x_i) = 
\left\{
\begin{array}{l@{}l}
\hat{u}_i(\gb) + \hat{v}_l(\gb) & \quad \text{if} \,\, \pi'_{i}(\gb) = l  \\
g_{l}(x_i) & \quad \text{otherwise}.
\end{array}
\right.
\end{equation}
Now, note that, for all $i \in \Ical$ and $l \in \Lcal$, it holds that $\hat{u}_i(\gb) + \hat{v}_l(\gb) \geq \hat{g}_{l}(x_i)$ 
and thus $\hat{\ub}(\gb)$ and $\hat{\vb}(\gb)$ is a feasible dual solution to the linear program
defined by Eq.~\ref{eq:matching-lp} under $\hat{\gb}$.
Moreover, for any $i \in \Ical$ and $l \in \Lcal$ such that $\hat{z}_{il}(\gb) = 1$, it holds that $\hat{u}_i(\gb) + \hat{v}_l(\gb) = \hat{g}_{l}(x_i)$ independently of the 
value of $\hat{z}_{il}(\hat{\gb})$.
As a direct consequence, we have that $\hat{\zb}(\gb)$ is {\em still} an optimal solution to the linear program defined by Eq.~\ref{eq:matching-lp} under $\hat{\gb}$. 
Moreover, from complementary slackness, $\hat{\ub}(\gb)$ and $\hat{\vb}(\gb)$ are also an optimal solution to the dual of the linear program under $\hat{\gb}$. 
Thus, by strong duality, we have that:
\begin{equation} \label{eq:strong-duality}
\sum_{i \in \Ical, l \in \Lcal} {\hat{g}_{l}(x_i) \, \hat{z}_{il}(\gb)}
= \sum_{i \in \Ical} {\hat{u}_i(\gb)} + \sum_{l \in \Lcal} {c_l \cdot \hat{v}_l(\gb)}
= \sum_{i \in \Ical, l \in \Lcal} {g_{l}(x_i) \, \hat{z}_{il}(\gb)}
\end{equation}
Next, let $\zb' = (z'_{il})_{i \in \Ical, l \in |Lcal}$ be defined as $z'_{il}=\mathds{1}[\pi'_i(\gb)=l]$.
Then, 
using Eqs.~\ref{eq:invlp_weights} and~\ref{eq:strong-duality} and the fact that $n = \sum_{l \in \Lcal} c_l$, we have 
that:
\begin{align*}
\sum_{i \in \Ical, l \in \Lcal} {\hat{g}_{l}(x_i) \, z'_{il}}
&= \sum_{i \in \Ical, l \in \Lcal} \left[ \mathds{1}[\pi'_{i}(\gb)=l] \left( \hat{u}_i(\gb) + \hat{v}_l(\gb) \right) z'_{il}(\gb) + \mathds{1}[\pi'_{i}(\gb)\neq l] \cdot g_{l}(x_i) z'_{il}(\gb) \right] \\
&= \sum_{i \in \Ical, l \in \Lcal} \mathds{1}[\pi'_{i}(\gb)=l] \left( \hat{u}_i(\gb) + \hat{v}_l(\gb) \right) z'_{il}(\gb) \\
&= \sum_{i \in \Ical} \hat{u}_i(\gb) \left[ \sum_{l \in \Lcal} \mathds{1}[\pi'_{i}(\gb)=l] \cdot z'_{il}\right] + \sum_{l \in \Lcal} \hat{v}_{l}(\gb) \left[ \sum_{i \in \Ical} \mathds{1}[\pi'_{i}(\gb)=l] \cdot z'_{il} \right] \\
&= \sum_{i \in \Ical} {\hat{u}_i(\gb)} + \sum_{l \in \Lcal} {c_l \cdot \hat{v}_l(\gb)} \\
&= \sum_{i \in \Ical, l \in \Lcal} {\hat{g}_{l}(x_i) \, \hat{z}_{il}(\gb)} \\
&= \sum_{i \in \Ical, l \in \Lcal} {g_{l}(x_i) \, \hat{z}_{il}(\gb)}.
\end{align*}
As a direct consequence, we have that $\zb'$ is an optimal solution to the linear program defined by Eq.~\ref{eq:matching-lp} under $\gb$ and thus also under $\hat{\gb}$. 
Moreover, using the definition of $\hat{\pi}(\gb)$, we also have that:
\begin{equation} \label{eq:equality-hatg-g}
\sum_{i \in \Ical} \hat{g}_{\pi'_{i}(\gb)}(x_i)
= \sum_{i \in \Ical, l \in \Lcal} {\hat{g}_{l}(x_i) \, z'_{il}}
= \sum_{i \in \Ical, l \in \Lcal} {g_{l}(x_i) \, \hat{z}_{il}(\gb)}
= \sum_{i \in \Ical} g_{\hat{\pi}_{i}(\gb)}(x_i)
\end{equation}
Further, using the definition of $\hat{\gb}$, the complementary slackness conditions and Eq.~\ref{eq:equality-hatg-g}, we have that:
\begin{equation*}
\sum_{i \in \Ical, l \in \Lcal} | \hat{g}_{l}(x_i) - g_{l}(x_i) |
= \sum_{i \in \Ical} \left[ \hat{g}_{\pi'_{i}(\gb)}(x_i) - g_{\pi'_{i}(\gb)}(x_i) \right]
= \sum_{i \in \Ical} \left[ g_{\hat{\pi}_{i}(\gb)}(x_i) - g_{\pi'_{i}(\gb)}(x_i) \right]
\end{equation*}
Since $\pi'(\gb) \in \hat{\Pi}(\hat{\gb})$, this directly implies that, for any $\hat{\pi}(\gb) \in \hat{\Pi}(\gb)$ and $\hat{\pi}(\hat{\gb}) \in \hat{\Pi}(\hat{\gb})$, it holds that
\begin{equation*}
\sum_{i \in \Ical, l \in \Lcal} | \hat{g}_{l}(x_i) - g_{l}(x_i) |
= \sum_{i \in \Ical} \left[ g_{\hat{\pi}_{i}(\gb)}(x_i) - g_{\hat{\pi}_{i}(\hat{\gb})}(x_i) \right]
\end{equation*}
Thus, from Lemma~\ref{lemma:lower-bound}, it follows that the predicted probabilities $\hat{\gb}$ must be the minimally modified predicted probabilities $\gb'$ under which $\pi'(\gb) \in \hat{\Pi}(\gb')$.
This concludes the proof.

%% file: 083synthetic.tex
In this section, we give additional details regarding the procedure we follow to generate synthetic refugee from aggregate statistics on demographic features of refugees and countries from various international organizations.

\subsection{Aggregate Statistical Data}
\label{sec:statistical-data}
We gather aggregate statistics about refugees who migrated to the United States in 2022 from the UNHCR database~\cite{unhcr2022refugee}, including country of origin, sex, age groups, and number of refugees allocated across ten states, namely California, Florida, Illinois, Maryland, Massachusetts, New Jersey, New York, Pennsylvania, Texas and Virginia. 
We gather aggregate statistics about the population of the above ten US states from the U.S. Census~\cite{census2022education}, including age, sex, and education. 
We gather aggregate statistics about employment in the above ten US states for different age groups and sex from the U.S. Department of Labor~\cite{bereau2024labor}.
We gather aggregate statistics about the foreign-born population in the above ten US states from the Migration Policy Institute~\cite{migrationpolicy2024profiles}, including sex, country of origin and employment.
Additionally, we gather aggregate statistics about the population of different countries from the World Bank~\cite{worldbank2024world}, including age, sex, and education.
We standardize statistics on age groups, country of origin, and education levels across data sources.
The age groups are \emph{20-24}, \emph{25-34}, \emph{35-44}, \emph{45-64}, and \emph{65-100}.
The levels of education are \emph{Primary or less}, \emph{Secondary}, and \emph{Tertiary}.
The country of origin is classified into the following regions: \emph{Africa}, \emph{Asia}, \emph{Europe}, \emph{Latin America}, \emph{Northern America}, and \emph{Oceania}.
Table~\ref{tab:data-symbols} summarizes all the quantities used in our data generation process, which are derived from the above aggregate statistics.

\begin{table}[h!]
\caption{Quantities used in our data generation process, which are derived from publicly available aggregate statistics. The symbol $w=1$ indicates that the given statistics pertain to the employed population, and $f=1$ indicates that the statistics pertain to the foreign-born population. If both $w=1$ and $f=1$ are present, the statistics refer to the employed foreign-born population.}
\resizebox{\textwidth}{!}
{
\begin{tabular}{@{}ll@{}}
\toprule
Symbol & \multicolumn{1}{c}{Meaning} \\ \midrule
$\tau(c, s)$ & The proportion of refugees with country of origin $c$ and sex $s$ \\
$\tau(a \given c)$ & The proportion of the population in age group $a$ in country of origin $c$ \\
$\tau(e \given c)$ & The proportion of the population with level of education $e$ in country of origin $c$ \\
$n(a, e, s \given l)$ & The number of people in state $l$ in age group $a$, with level of educational $e$, and sex $s$ \\
$\tau(a \given l, w=1)$ & The proportion of the employed population in state $l$ in age group $a$ \\
$\tau(s \given l, w=1)$ & The proportion of the employed population in state $l$ of sex $s$\\
$\tau(w=1 \given l, f=1)$ & The proportion of the foreign-born population in state $l$ employed.\\
$\tau(c \given l, f=1)$ & The proportion of the foreign-born population in state $l$ from country of origin $c$ \\
$\tau(c \given l, w=1, f=1)$ & The proportion of the employed, foreign-born population in state $l$ from country of origin $c$ \\
$\tau(e \given l, f=1)$ & The proportion of the foreign-born population in state $l$ with level of education $e$\\
$\tau(e \given l, w=1, f=1)$ & The proportion of the employed, foreign-born population in state $l$ with level of education $e$ \\
$\tau(s \given l, f=1)$ & The proportion of the foreign-born population in state $l$ of sex $s$ \\ \bottomrule
\end{tabular}
}
\label{tab:data-symbols}
\vspace{-3mm}
\end{table}

\subsection{Data Generation Process}
\label{sec:data-synthesis}
The attributes of each refugee $i$ --- age group $a_i \in \Acal$, country of origin $c_i \in \Ccal$, level of education $e_i \in \Ecal$ and sex $s_i \in \Scal$---are drawn according to different categorical distributions. First, we draw the country of origin $c_i$ and the sex $s_i$ based on the joint categorical distribution parameterized by proportions $\tau(c, s)$. The age group $A=a_i$ and level of education $E=e_i$ are then drawn per individual depending on their country of origin $c_i$:
\begin{equation}
   c_{i},s_{i} \sim Cat(\tau(c, s)), \quad a_{i} \sim Cat(\tau(a\given c_{i})), \quad e_{i}\sim Cat(\tau(e\given c_{i}))
\end{equation}
Employment probability $p_{i,l}$ of refugee $i$ in location $l$ follows a beta distribution $\Bcal_i$ with a mean of $\mu_{l}(a_i, c_i, e_i, s_i)$---the average employment probability of refugees with features $a_i$, $c_i$, $e_i$, $s_i$ in state $l$---and a fixed variance $\sigma^2$.
The variance $\sigma^2$ is set to $0.001$ and the mean $\mu_{l}(a,c,e,s)$ is chosen to match the marginal statistics $\mu_{l}(a)$, $\mu_{l}(c)$, $\mu_{l}(e)$ and $\mu_{l}(s)$ of the general population in the US by solving the following optimization problem: 
\begin{align}\label{eq:quad_programm}
\begin{split}
    \min_{\{\mu_{l}(a,c,e,s)\}
    } \quad & { \frac{1}{|\Ccal \times \Ecal\times \Scal|} \sum_{c',e',s'} (\mu_{l}(a,c',e',s') - \mu_{l}(a))^2 } + \frac{1}{|\Acal \times \Ecal \times \Scal|}\sum_{a',e',s'} (\mu_{l}(a',c,e',s') - \mu_{l}(c))^2 \\
    &+ \frac{1}{|\Acal \times \Ccal \times \Scal|}\sum_{a',c',s'} (\mu_{l}(a',c',e,s') - \mu_{l}(e))^2 + \frac{1}{|\Acal \times \Ccal \times \Ecal|}\sum_{a',c',e'} (\mu_{l}(a',c',e',s) - \mu_{l}(s))^2\\
    \text{subject to} \quad
    &\mu_{l}(a) (1 - \rho_a) \leq \sum_{c',e',s'}\mu_{l}(a,c',e',s') \cdot \tau(c', e', s' \given l, f=1) 
    \leq \mu_{l}(a) (1 + \rho_a) \\
    &\mu_{l}(c) (1 - \rho_c) \leq \sum_{a',e',s'}\mu_{l}(a',c,e',s') \cdot \tau(a', e', s' \given l, f=1) 
    \leq \mu_{l}(c) (1 + \rho_c) \\
    & \mu_{l}(e) (1 - \rho_e) \leq \sum_{a',c',s'}\mu_{l}(a',c',e,s') \cdot \tau(a',c',s' \given l, f=1) 
    \leq \mu_{l}(e) (1 + \rho_e) \\
    & \mu_{l}(s) (1 - \rho_s) \leq \sum_{a',c',e'}\mu_{l}(a',c',e',s) \cdot \tau(a',c',e' \given l, f=1) 
    \leq \mu_{l}(s) (1 + \rho_s) \\
    & \mu^L_l(a, c, e, s) (1 - \rho_b) \leq \mu_{l}(a,c,e,s) \leq \mu^{U}_l(a, c, e, s) (1 + \rho_b).
\end{split}
\end{align}
Here, $\tau(a, c, e \given l, f=1)$ denotes the proportion of the foreign-born population in age group $a$, from country of origin $c$, and with level of education $e$ in location $l$. The proportions $\tau(a, c, s \given l, f=1)$, $\tau(a, e, s \given l, f=1)$, and $\tau(c, e, s \given l, f=1)$ are defined analogously.
Under some assumptions, we can derive the above proportions from the aggregate statistics we gathered:
\begin{itemize}
    \item $\tau(c, e, s \given l, f=1) = \tau(c \given l, f=1) \cdot \tau(e, s \given l)$
    \item $\tau(a, e, s \given l, f=1) = \tau(a, e, s \given l)$
    \item $\tau(a, c, s \given l, f=1) = \tau(c \given l, f=1) \cdot \tau(a, s \given l)$
    \item $\tau(a, c, e \given l, f=1) = \tau(c \given l, f=1) \cdot \tau(a, e \given l)$
\end{itemize}
The first assumption is that the joint distribution of the employed population in age group $a$, with level of education $e$, and sex $s$ would be similar to that of the overall population with the same features across the states. The second assumption is that the country of origin $c$ is independent of $a$, $e$ and $s$ for the employed population across the states.

\begin{figure}[t]
    \centering
    \subfloat[California]{
         \includegraphics[width=.19\linewidth]{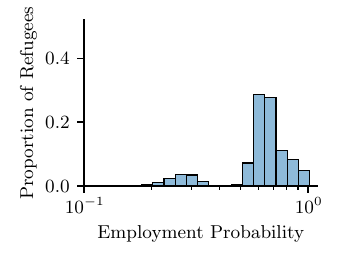}
    }
    \subfloat[Florida]{
         \includegraphics[width=.19\linewidth]{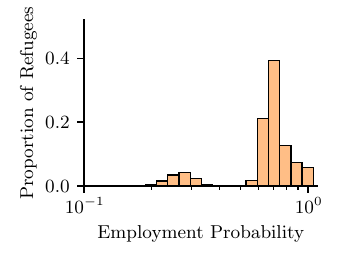}
    }
    \subfloat[Illinois]{
         \includegraphics[width=.19\linewidth]{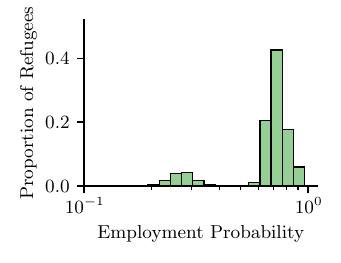}
    }
    \subfloat[Maryland]{
         \includegraphics[width=.19\linewidth]{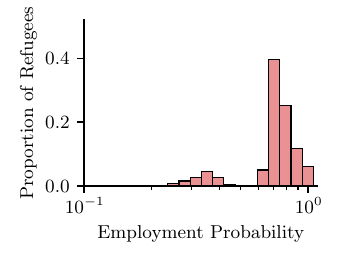}
    }
    \subfloat[Massachusetts]{
         \includegraphics[width=.19\linewidth]{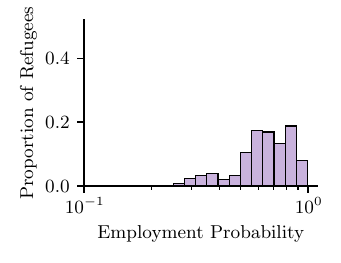}
    }
    \\
    \subfloat[New Jersey]{
         \includegraphics[width=.19\linewidth]{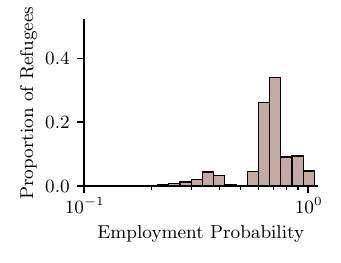}
    }
    \subfloat[New York]{
         \includegraphics[width=.19\linewidth]{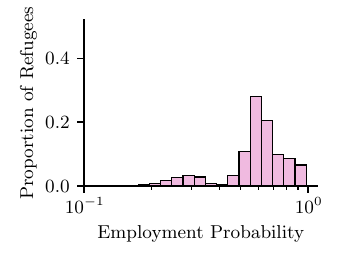}
    }
    \subfloat[Pennsylvania]{
         \includegraphics[width=.19\linewidth]{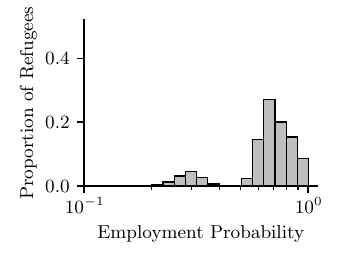}
    }
    \subfloat[Texas]{
         \includegraphics[width=.19\linewidth]{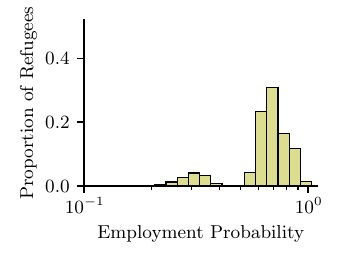}
    }
    \subfloat[Virginia]{
         \includegraphics[width=.19\linewidth]{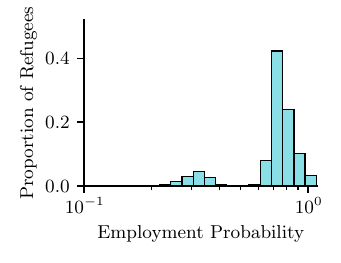}
    }
    
    \caption{Empirical distribution of employment probability of refugees for all states, computed using $500{,}000$ synthesized refugees. The distributions are plotted using exponential binning.}
    \label{fig:distribution-employment-probability}
    \vspace{-3mm}
\end{figure}

\begin{figure}[t]
    \centering
    \subfloat[California vs Virginia]{
         \includegraphics[width=.35\linewidth]{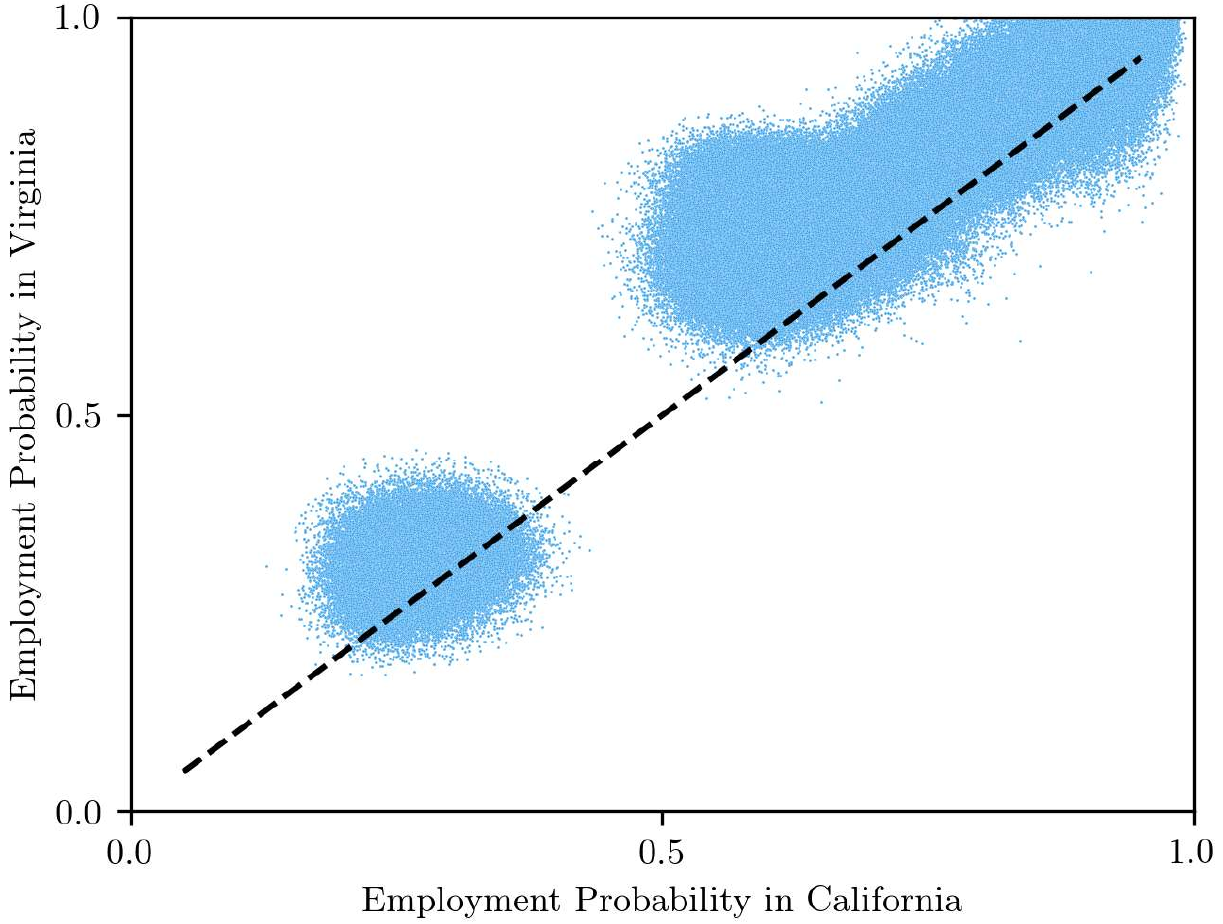}
    }
    \hspace{5mm}
    \subfloat[Florida vs Massachusetts]{
         \includegraphics[width=.35\linewidth]{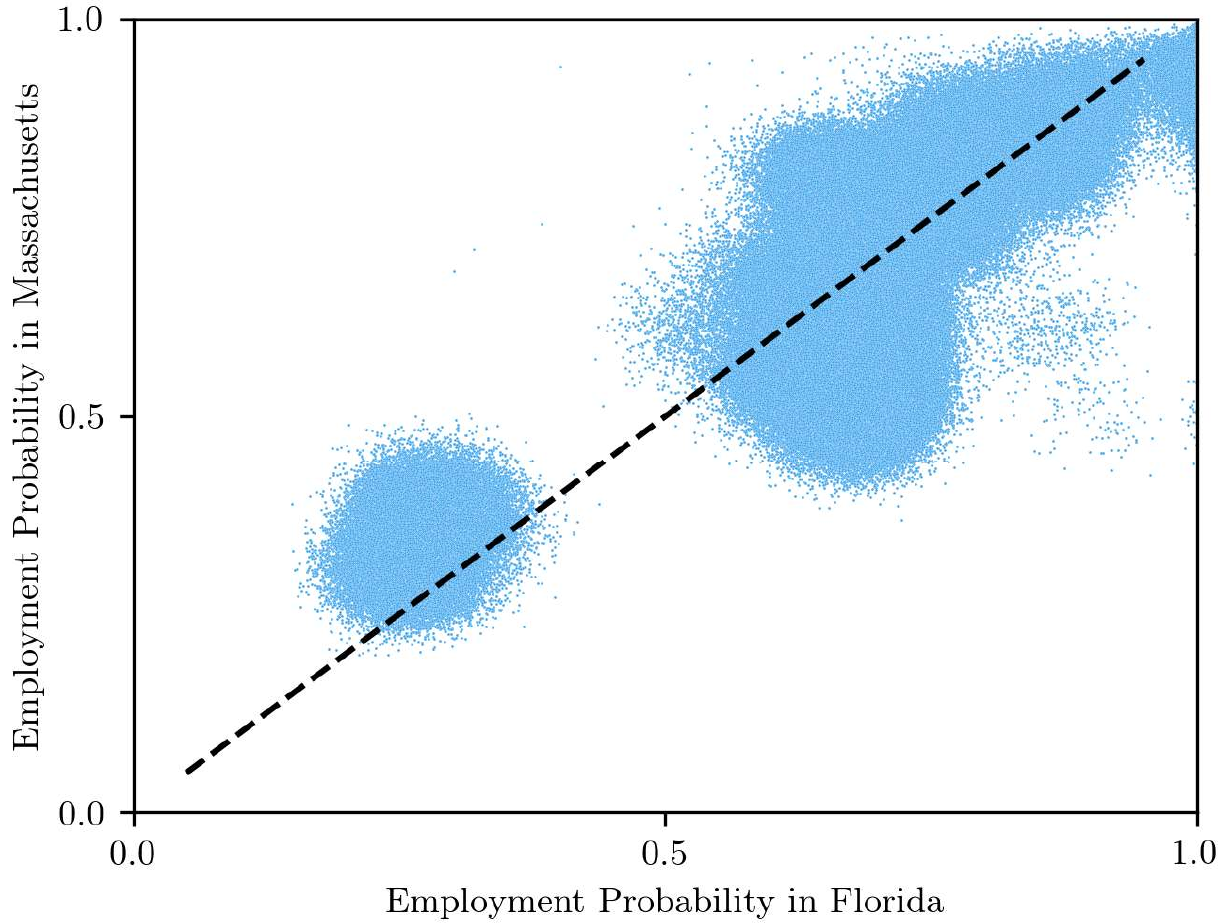}
    }
    
    \caption{Per-refugee employment probability for two pairs of locations and $500,000$ synthesized refugees. The (dashed) identity line (\ie, $y=x$) indicates equal probabilities between two states.}
    \label{fig:employment-probability-comparison}
    \vspace{-3mm}
\end{figure}

In Eq.~\ref{eq:quad_programm}, the hyperparameters $\rho_a$, $\rho_c$, $\rho_e$, $\rho_s$ control the ``looseness'' of the conditions that make $\mu_l(a, c, e, s)$ follow the marginal statistics $\mu_l(a)$, $\mu_l(c)$, $\mu_l(e)$, $\mu_l(s)$, respectively, based on the proportion of each refugee group. We set $\rho_a=0.5$, $\rho_c=0.0$, $\rho_e=0.1$, and $\rho_s=0.0$. 
The possible range of the solution is decided by another hyperparameter $\rho_b$ as well as $\mu^L_{l}(a,c,e,s) = \max (\mu_l(a), \mu_l(c), \mu_l(e), \mu_l(s))$ and $\mu^U_{l}(a,c,e,s) = \min (\mu_l(a), \mu_l(c), \mu_l(e), \mu_l(s))$.
This prevents the solver finding an extreme solution for $\mu_{l}(a,c,e,s)$ that is far from marginal statistics.
In our generation process $\rho_b$ is set to $0.6$.

Marginal statistics $\mu_{l}(a)$, $\mu_{l}(c)$, $\mu_{l}(e)$ and $\mu_{l}(s)$ are computed based on the available aggregate statistics as follows.
Marginal statistics $\mu_{l}(a)$ (\ie, the proportion of the foreign-born population in state $l$ in age group $a$ employed) is computed as follows:
\begin{equation*}
\mu_{l}(a) =\tau(w=1 | a,l,f=1) = \frac{\tau(a \given l, w=1, f=1) \cdot \tau(w=1 \given l, f=1)}{\tau(a \given l, f=1)} = \frac{\tau(a \given l, w=1) \cdot \tau(w=1 \given l, f=1)}{\tau(a \given l)},
\end{equation*}
where $\tau(a \given l) = \frac{\sum_{e' \in \Ecal, s' \in \Scal}{n(a, e', s' \given l)}}{\sum_{a' \in \Acal, e' \in \Ecal, s' \in \Scal}{n(a', e', s' \given l)}}$. Here, we assume that the age distribution of the employed, foreign-born population would be similar to that of total employed population across the states (\ie, $\tau(a \given l, w=1, f=1) \approx \tau(a \given l, w=1)$), and the age distribution of the foreign-born population would be similar to that of the total population across the states (\ie, $\tau(a \given l, f=1) \approx  \tau(a \given l)$).
Marginal statistics $\mu_{l}(c)$ (\ie, the proportion of the foreign-born population in state $l$ from country of origin $c$ employed ) is computed as follows:
\begin{equation*}
\mu_{l}(c) = \tau(w=1 | c,l,f=1) = \frac{\tau(c \given l, w=1, f=1) \cdot \tau(w=1 \given l, f=1)}{\tau(c \given l, f=1)}.
\end{equation*}
Marginal statistics $\mu_{l}(e)$ (\ie, the proportion of the foreign-born population in state $l$ with level of education $e$ employed) is computed as follows:
\begin{equation*}
\mu_{l}(e) 
=\tau(w=1 | e,l,f=1) = \frac{\tau(e \given l, w=1, f=1) \cdot \tau(w=1 \given l, f=1)}{\tau(e \given l, f=1)}.
\end{equation*}
Marginal statistics $\mu_{l}(s)$ (\ie, the proportion of the foreign-born population in state $l$ of sex $s$ employed) is computed as follows:
\begin{equation*}
\mu_{l}(g) 
=\tau(w=1 | s,l,f=1) = \frac{\tau(s \given l, w=1, f=1) \cdot \tau(w=1 \given l, f=1)}{\tau(s \given l, f=1)} =\frac{\tau(s \given l, w=1) \cdot \tau(w=1 \given l, f=1)}{\tau(s \given l, f=1)},
\end{equation*}
where we assume that the sex distribution of the employed, foreign-born population would be similar to that of total employed population across the states (\ie, $\tau(s \given l, w=1, f=1) = \tau(s \given l, w=1)$).

\subsection{Distribution of Generated Data}
\label{sec:data-distribution}
Using the generation process described in Appendix~\ref{sec:data-synthesis}, we create $5{,}000$ synthetic pools of refugees to be resettled to $k=10$ locations.
Each pool contains $n=100$ synthetic refugees and the employment probability of each refugee is sampled from the beta distribution derived from their demographic features.
Figure~\ref{fig:distribution-employment-probability} shows the empirical distribution of employment probability for all states, computed using the resulting $500{,}000$ synthesized refugees.
%
%
While these empirical distributions are similar, the employment probability per refugee varies across locations, as shown in Figure~\ref{fig:employment-probability-comparison} for two pairs of states.

%% file: 084implementation.tex
In this section, we offer additional implementation details about deep learning model $h$ used in our postprocessing framework.
\medskip

\xhdr{Training Data} 
The minimally modified employment probabilities $\breve{g}$ are computed from the given placement decisions made by the default policy $\tilde{\pi}(\xb, w)$ and the predicted employment probabilities $\gb$ by solving Eq.~\ref{eq:partial-dual} using the linear program solver in {\tt Scipy} package.

\xhdr{Architecture of the Postprocessing Deep Learning Model $h$}
Our deep learning model $h$ consists of a prediction probability projection layer, a capacity projection layer, $N$ layers of Transformer encoders, and an embedding projection layer.
Note that all projection layers consist of two linear projections with ReLU activation.
A prediction probability projection layer projects $\gb$, the expected employment probability of each refugee in $k=10$ locations, to hidden dimension $d=128$.
Similarly, a capacity projection layer projects the capacity of all locations to the same hidden dimension. This projected capacity information is then added to the projected employment probabilities for each refugee.
The added data pass through $N=2$ layers of Transformer encoder, where each Transformer encoder layer has $1$ head and $128$ hidden dimensions.
Note that, the layers of Transformer encoder exclude positional encoding so that they become agnostic to the number or order of the refugees in each pool.
The processed data is projected onto a $k=10$ dimensional space to predict the difference $\breve{\gb} - \gb$. The final output of the model $h(\gb)$ is obtained by adding this predicted difference to the original model input $\gb$.

\medskip
\xhdr{Training}
In all our experiments, we employ quadratic loss and the AdamW optimizer for the optimization process. Our training batches consist of $16$ pools of refugees, and the training duration is set for $50$ epochs. We start with an initial learning rate of $0.001$ and apply an exponential learning rate scheduler with a decay factor of $\gamma=0.9$ to adjust the learning rate over time.
It takes $7$ seconds of training time per epoch in our setting.

\medskip
\xhdr{Hardware Setup}
We use a GPU server equipped with Intel Xeon Platinum 8268 CPU @ 2.90GHz, 376 GB memory and NVIDIA A100 GPUs. A single GPU is used in each experiment.

%% file: 085additional_experiments.tex
In this section, we provide additional experimental results that are omitted from the main paper due to space limitations.

\subsection{Experimental results on realized utility}
In Section~\ref{sec:experiments}, we could verify that the algorithmic policy $\hat{\pi}(\breve{\gb})$  reduces the risk of harm compared to algorithmic policy $\hat{\pi}(\gb)$.
Table~\ref{tab:full_empirical_utility} summarizes the per-pool average realized utility of the algorithmic policies given labels $\{y_i(l)\}_{l\in\Lcal, i\in \Ical}$ indicating whether a refugee would find employment at each state $l$ soon after relocation.
As expected from Eq.~\ref{eq:harmless-average}, we observe that the unreali\-za\-ble algorithmic policy $\hat{\pi}(\breve{\gb})$ achieves higher average realized utility compared to default policy $\tilde{\pi}(\xb, w)$ across all noise levels. 
Additionally, the postprocessing algorithmic policy $\hat{\pi}(h(\gb))$ successfully achieves higher average realized utility compared to the algorithmic policy $\hat{\pi}(\gb)$ that maximizes the predicted utility under the predicted employment probabilities $\gb$.
\begin{table}[h]
\centering
\caption{Comparison of the per-pool average realized utility of algorithmic policies $\tilde{\pi}(\xb, w)$, $\hat{\pi}(\pb)$, $\hat{\pi}(\gb)$, $\hat{\pi}(\breve{\gb})$, and $\hat{\pi}(h(\gb))$ against the default policy $\tilde{\pi}(\xb, w)$ across 500 test set pools under various noise levels $w$ (higher numbers indicate better performance). 
The numbers of $\hat{\pi}(h(\gb))$ are averaged results over 5 runs with their standard deviation.
Bold numbers indicate increase in average realized utility by $\hat{\pi}(h(\gb))$ compared to $\hat{\pi}(\gb)$, while underlined numbers indicates the higher average realized utility of $\hat{\pi}(h(\gb))$ compared to $\tilde{\pi}(\xb, w)$.
}
\begin{tabular}{@{}c|ccccc@{}}
\toprule
Noise Level ($w$) & $\tilde{\pi}(\xb, w)$ & $\hat{\pi}(\pb)$ & $\hat{\pi}(\gb)$ & $\hat{\pi}(\breve{\gb})$ & $\hat{\pi}(h(\gb))$ \\ \midrule
0.000 & 0.7200 & 0.7200 & 0.6976 & 0.8161 & \textbf{0.7077$\pm$0.0009} \\
0.125 & 0.7129 & 0.7200 & 0.6976 & 0.8180 & \textbf{0.7074$\pm$0.0003} \\
0.250 & 0.7032 & 0.7200 & 0.6976 & 0.8188 & {\underline{\textbf{0.7072$\pm$0.0012}}} \\
0.375 & 0.6956 & 0.7200 & 0.6976 & 0.8195 & {\underline{\textbf{0.7071$\pm$0.0015}}} \\
0.500 & 0.6892 & 0.7200 & 0.6976 & 0.8242 & {\underline{\textbf{0.7038$\pm$0.0012}}} \\
0.625 & 0.6794 & 0.7200 & 0.6976 & 0.8262 & {\underline{\textbf{0.7007$\pm$0.0012}}} \\
0.750 & 0.6715 & 0.7200 & 0.6976 & 0.8288 & {\underline{\textbf{0.7002$\pm$0.0006}}} \\
0.875 & 0.6659 & 0.7200 & 0.6976 & 0.8293 & {\underline{\textbf{0.6981$\pm$0.0005}}} \\
1.000 & 0.6532 & 0.7200 & 0.6976 & 0.8319 & {\underline{0.6975$\pm$0.0005}} \\ \bottomrule
\end{tabular}
\label{tab:full_empirical_utility}
\end{table}

\clearpage

\subsection{Full Experimental results on counterfactual utility}
Table~\ref{tab:full_harm_pool_ratio} presents the full results for the percentage (\%) of pools in the test that are counterfactually harmed by each algorithmic policy under different noise levels $w$, and Table~\ref{tab:full_counterfactual_utility} presents the full results for the per-pool expected counterfactual utilities of each algorithmic policy under different noise levels $w$. The numbers of $\hat{\pi}(h(\gb))$ are average results over 5 runs with their standard deviation. In both tables, we can observe that the algorithmic policy that uses our postprocessing algorithm, $\hat{\pi}(h(\gb))$, reduces (maintains) harm in high (low) noise levels compared to the algorithmic policy that uses the predicted employment probability, $\hat{\pi}(\gb)$.
In Table~\ref{tab:full_harm_pool_ratio}, the percentage of pools counterfactually harmed by $\hat{\pi}(\breve{\gb})$ is $0\%$ for all noise level, as expected from Eq.~\ref{eq:harmless-average}.
In Table~\ref{tab:full_counterfactual_utility}, $\hat{\pi}(\breve{\gb})$ consistently exhibits higher expected counterfactual utility across all noise levels. Furthermore, $\hat{\pi}(h(\gb))$ shows higher expected counterfactual utility compared to the default policy $\tilde{\pi}(\xb, w)$ for high noise levels (\ie, $w \geq 0.25$).
\begin{table}[h]
\centering
\caption{
Comparison of the percentage (\%) of pools counterfactually harmed by algorithmic policies $\hat{\pi}(\pb)$, $\hat{\pi}(\gb)$, $\hat{\pi}(\breve{\gb})$, and $\hat{\pi}(h(\gb))$ against the default policy $\tilde{\pi}(\xb, w)$ across 500 test set pools under various noise levels $w$ (lower numbers indicate better performance). 
The numbers of $\hat{\pi}(h(\gb))$ are averaged results over 5 runs with their standard deviation.
Bold numbers indicate reductions in counterfactually harmed pools by $\hat{\pi}(h(\gb))$ compared to $\hat{\pi}(\gb)$.
}
\vspace{1mm}
\begin{tabular}{@{}c|cccc@{}}
\toprule
Noise Level ($w$) & $\hat{\pi}(\pb)$ & $\hat{\pi}(\gb)$ & $\hat{\pi}(\breve{\gb})$ & $\hat{\pi}(h(\gb))$ \\ \midrule
0.000 & 0.00 & 75.00 & 0.00 & \textbf{69.24$\pm$0.96} \\
0.125 & 33.20 & 64.60 & 0.00 & \textbf{58.88$\pm$0.63} \\
0.250 & 23.60 & 55.00 & 0.00 & \textbf{45.44$\pm$0.59} \\
0.375 & 17.60 & 48.20 & 0.00 & \textbf{38.00$\pm$1.25} \\
0.500 & 13.60 & 41.00 & 0.00 & \textbf{33.92$\pm$0.68} \\
0.625 & 11.40 & 33.40 & 0.00 & \textbf{30.64$\pm$1.04} \\
0.750 & 9.20 & 24.80 & 0.00 & \textbf{22.88$\pm$0.56} \\
0.875 & 8.20 & 23.40 & 0.00 & 23.40$\pm$0.77 \\
1.000 & 6.00 & 15.60 & 0.00 & 16.04$\pm$0.50 \\ \bottomrule
\end{tabular}
\label{tab:full_harm_pool_ratio}
\end{table}

\begin{table}[h]
\caption{
Comparison of the expected counterfactual utility of algorithmic policies $\tilde{\pi}(\xb, w)$, $\hat{\pi}(\pb)$, $\hat{\pi}(\gb)$, $\hat{\pi}(\breve{\gb})$, and $\hat{\pi}(h(\gb))$ against the default policy $\tilde{\pi}(\xb, w)$ across 500 test set pools under various noise levels $w$ (higher numbers indicate better performance). 
The numbers of $\hat{\pi}(h(\gb))$ are averaged results over 5 runs with their standard deviation.
Bold numbers indicate increase in expected counterfactual utility by $\hat{\pi}(h(\gb))$ compared to $\hat{\pi}(\gb)$, while underlined numbers indicates the higher expected counterfactual utility of $\hat{\pi}(h(\gb))$ compared to $\tilde{\pi}(\xb, w)$.
}
\vspace{1mm}
\centering
\begin{tabular}{@{}c|ccccc@{}}
\toprule
Noise Level ($w$) & $\tilde{\pi}(\xb, w)$ & $\hat{\pi}(\pb)$ & $\hat{\pi}(\gb)$ & $\hat{\pi}(\breve{\gb})$ & $\hat{\pi}(h(\gb))$ \\ \midrule
0.000 & 0.7200 & 0.7200 & 0.6991 & 0.8156 & \textbf{0.7078$\pm$0.0004} \\
0.125 & 0.7129 & 0.7201 & 0.6989 & 0.8179 & \textbf{0.7072$\pm$0.0003} \\
0.250 & 0.7032 & 0.7192 & 0.6976 & 0.8192 & {\underline{\textbf{0.7074$\pm$0.0003}}} \\
0.375 & 0.6956 & 0.7202 & 0.6984 & 0.8220 & {\underline{\textbf{0.7064$\pm$0.0008}}} \\
0.500 & 0.6892 & 0.7210 & 0.6974 & 0.8244 & {\underline{\textbf{0.7044$\pm$0.0005}}} \\
0.625 & 0.6794 & 0.7180 & 0.6967 & 0.8248 & {\underline{\textbf{0.7005$\pm$0.0008}}} \\
0.750 & 0.6715 & 0.7201 & 0.6982 & 0.8290 & {\underline{\textbf{0.7001$\pm$0.0004}}} \\
0.875 & 0.6659 & 0.7187 & 0.6964 & 0.8298 & {\underline{\textbf{0.6969$\pm$0.0001}}} \\
1.000 & 0.6532 & 0.7184 & 0.6967 & 0.8309 & {\underline{0.6965$\pm$0.0001}} \\ \bottomrule
\end{tabular}
\label{tab:full_counterfactual_utility}
\end{table}

\clearpage

\subsection{Full Experimental Results of Hyperparameter Analysis}

Table~\ref{tab:full-utility-beta} shows the full results of expected counterfactual utility achieved by the algorithmic policies $\hat{\pi}(\pb)$, $\hat{\pi}(\gb)$, $\hat{\pi}(\breve{\gb})$ and $\hat{\pi}(h(\gb))$ in comparison with the expected realized utility achieved by the default policy $\tilde{\pi}(\xb, w)$ for different $\beta$ values under varying noise levels. 
We can observe that the postprocessing algorithmic policy offers greater expected counterfactual utility than $\hat{\pi}(\gb)$ in most of the cases.
\begin{table}[h]
\centering
\caption{
Expected counterfactual utility achieved by the algorithmic policies $\hat{\pi}(\pb)$, $\hat{\pi}(\gb)$ and $\hat{\pi}(h(\gb))$ in comparison with the expected realized utility achieved by the default policy $\tilde{\pi}(\xb, w)$ across all pools in the test set for different $\beta$ values under varying noise levels.
For $\hat{\pi}(h(\gb))$, the results are averaged over $5$ runs. 
}
\resizebox{\textwidth}{!}
{
\begin{tabular}{@{}lcccccccccc@{}}
\toprule
\multicolumn{1}{c}{$\beta$} & Noise Level ($w$) & 0.000 & 0.125 & 0.250 & 0.375 & 0.500 & 0.625 & 0.750 & 0.875 & 1.000 \\ \midrule
 & $\tilde{\pi}(\xb, w)$ & 0.7200 & 0.7129 & 0.7032 & 0.6956 & 0.6892 & 0.6794 & 0.6715 & 0.6659 & 0.6532 \\
 & $\hat{\pi}(\pb)$ & 0.7200 & 0.7201 & 0.7192 & 0.7202 & 0.7210 & 0.7180 & 0.7201 & 0.7187 & 0.7184 \\ \midrule
\multicolumn{1}{c}{0.1} & $\hat{\pi}(\gb)$ & 0.7168 & 0.7166 & 0.7157 & 0.7169 & 0.7169 & 0.7155 & 0.7163 & 0.7145 & 0.7149 \\
 & $\hat{\pi}(\breve{\gb})$ & 0.7558 & 0.7686 & 0.7803 & 0.7913 & 0.8032 & 0.8106 & 0.8223 & 0.8306 & 0.8389 \\
 & $\hat{\pi}(h(\gb))$ & \textbf{0.7197} & {\underline{\textbf{0.7194}}} & {\underline{\textbf{0.7192}}} & {\underline{\textbf{0.7196}}} & {\underline{\textbf{0.7199}}} & {\underline{\textbf{0.7172}}} & {\underline{\textbf{0.7170}}} & {\underline{\textbf{0.7150}}} & {\underline{\textbf{0.7153}}} \\ \midrule
\multicolumn{1}{c}{0.2} & $\hat{\pi}(\gb)$ & 0.7146 & 0.7145 & 0.7137 & 0.7144 & 0.7147 & 0.7135 & 0.7135 & 0.7117 & 0.7126 \\
 & $\hat{\pi}(\breve{\gb})$ & 0.7797 & 0.7867 & 0.7946 & 0.8018 & 0.8101 & 0.8163 & 0.8242 & 0.8299 & 0.8371 \\
 & $\hat{\pi}(h(\gb))$ & \textbf{0.7188} & {\underline{\textbf{0.7190}}} & {\underline{\textbf{0.7183}}} & {\underline{\textbf{0.7191}}} & {\underline{\textbf{0.7193}}} & {\underline{\textbf{0.7163}}} & {\underline{\textbf{0.7146}}} & {\underline{\textbf{0.7118}}} & {\underline{0.7126}} \\ \midrule
\multicolumn{1}{c}{0.3} & $\hat{\pi}(\gb)$ & 0.7125 & 0.7122 & 0.7117 & 0.7125 & 0.7130 & 0.7116 & 0.7121 & 0.7098 & 0.7103 \\
 & $\hat{\pi}(\breve{\gb})$ & 0.7905 & 0.7953 & 0.8018 & 0.8084 & 0.8140 & 0.8192 & 0.8258 & 0.8300 & 0.8356 \\
 & $\hat{\pi}(h(\gb))$ & \textbf{0.7170} & {\underline{\textbf{0.7169}}} & {\underline{\textbf{0.7163}}} & {\underline{\textbf{0.7165}}} & {\underline{\textbf{0.7169}}} & {\underline{\textbf{0.7132}}} & {\underline{\textbf{0.7123}}} & {\underline{\textbf{0.7099}}} & {\underline{0.7103}} \\ \midrule
\multicolumn{1}{c}{0.4} & $\hat{\pi}(\gb)$ & 0.7106 & 0.7104 & 0.7086 & 0.7105 & 0.7098 & 0.7089 & 0.7090 & 0.7081 & 0.7084 \\
 & $\hat{\pi}(\breve{\gb})$ & 0.7984 & 0.8029 & 0.8070 & 0.8122 & 0.8180 & 0.8218 & 0.8276 & 0.8300 & 0.8350 \\
 & $\hat{\pi}(h(\gb))$ & \textbf{0.7140} & \textbf{0.7144} & {\underline{\textbf{0.7133}}} & {\underline{\textbf{0.7129}}} & {\underline{\textbf{0.7127}}} & {\underline{\textbf{0.7092}}} & {\underline{\textbf{0.7094}}} & {\underline{\textbf{0.7082}}} & {\underline{\textbf{0.7086}}} \\ \midrule
\multicolumn{1}{c}{0.5} & $\hat{\pi}(\gb)$ & 0.7050 & 0.7046 & 0.7037 & 0.7048 & 0.7044 & 0.7032 & 0.7048 & 0.7029 & 0.7026 \\
 & $\hat{\pi}(\breve{\gb})$ & 0.8069 & 0.8098 & 0.8129 & 0.8165 & 0.8207 & 0.8230 & 0.8283 & 0.8302 & 0.8335 \\
 & $\hat{\pi}(h(\gb))$ & \textbf{0.7108} & \textbf{0.7104} & {\underline{\textbf{0.7096}}} & {\underline{\textbf{0.7098}}} & {\underline{\textbf{0.7090}}} & {\underline{\textbf{0.7055}}} & {\underline{\textbf{0.7050}}} & {\underline{0.7029}} & {\underline{\textbf{0.7028}}} \\ \midrule
\multicolumn{1}{c}{0.6} & $\hat{\pi}(\gb)$ & 0.6991 & 0.6989 & 0.6976 & 0.6984 & 0.6974 & 0.6967 & 0.6982 & 0.6964 & 0.6967 \\
 & $\hat{\pi}(\breve{\gb})$ & 0.8156 & 0.8179 & 0.8192 & 0.8220 & 0.8244 & 0.8248 & 0.8290 & 0.8298 & 0.8309 \\
 & $\hat{\pi}(h(\gb))$ & \textbf{0.7078} & \textbf{0.7072} & {\underline{\textbf{0.7074}}} & {\underline{\textbf{0.7064}}} & {\underline{\textbf{0.7044}}} & {\underline{\textbf{0.7005}}} & {\underline{\textbf{0.7001}}} & {\underline{\textbf{0.6969}}} & {\underline{0.6965}} \\ \midrule
\multicolumn{1}{c}{0.7} & $\hat{\pi}(\gb)$ & 0.6914 & 0.6908 & 0.6901 & 0.6908 & 0.6913 & 0.6896 & 0.6906 & 0.6887 & 0.6899 \\
 & $\hat{\pi}(\breve{\gb})$ & 0.8240 & 0.8253 & 0.8247 & 0.8269 & 0.8282 & 0.8270 & 0.8292 & 0.8288 & 0.8284 \\
 & $\hat{\pi}(h(\gb))$ & \textbf{0.7050} & \textbf{0.7041} & {\underline{\textbf{0.7037}}} & {\underline{\textbf{0.7021}}} & {\underline{\textbf{0.7004}}} & {\underline{\textbf{0.6974}}} & {\underline{\textbf{0.6960}}} & {\underline{\textbf{0.6898}}} & {\underline{0.6892}} \\ \midrule
\multicolumn{1}{c}{0.8} & $\hat{\pi}(\gb)$ & 0.6862 & 0.6857 & 0.6850 & 0.6852 & 0.6860 & 0.6839 & 0.6853 & 0.6832 & 0.6839 \\
 & $\hat{\pi}(\breve{\gb})$ & 0.8297 & 0.8292 & 0.8284 & 0.8290 & 0.8295 & 0.8275 & 0.8289 & 0.8280 & 0.8266 \\
 & $\hat{\pi}(h(\gb))$ & \textbf{0.7025} & \textbf{0.7020} & \textbf{0.7008} & {\underline{\textbf{0.6990}}} & {\underline{\textbf{0.6976}}} & {\underline{\textbf{0.6951}}} & {\underline{\textbf{0.6925}}} & {\underline{\textbf{0.6859}}} & {\underline{\textbf{0.6846}}} \\ \midrule
\multicolumn{1}{c}{0.9} & $\hat{\pi}(\gb)$ & 0.6839 & 0.6835 & 0.6827 & 0.6827 & 0.6835 & 0.6814 & 0.6827 & 0.6808 & 0.6817 \\
 & $\hat{\pi}(\breve{\gb})$ & 0.8309 & 0.8302 & 0.8290 & 0.8295 & 0.8296 & 0.8270 & 0.8281 & 0.8275 & 0.8257 \\
 & $\hat{\pi}(h(\gb))$ & \textbf{0.6996} & \textbf{0.6988} & \textbf{0.6972} & {\underline{\textbf{0.6969}}} & {\underline{\textbf{0.6954}}} & {\underline{\textbf{0.6888}}} & {\underline{\textbf{0.6852}}} & {\underline{\textbf{0.6826}}} & {\underline{0.6816}} \\ \bottomrule
\end{tabular}
}
\label{tab:full-utility-beta}
\end{table}